\documentclass[a4paper,onecolumn,11pt,accepted=2022-01-12]{quantumarticle}
\pdfoutput=1
\usepackage{graphicx, color}% Include figure files
\usepackage{dcolumn}% Align table columns on decimal point
\usepackage{bm}% bold math
\usepackage[utf8]{inputenc}
\usepackage{amsthm}
\usepackage{amsmath}
\usepackage{cite}
\usepackage[export]{adjustbox}
\newtheorem{thm}{Theorem}
\newtheorem{dfn}[thm]{Problem}

\usepackage{braket}
\usepackage{txfonts}
\usepackage[margin=15mm]{geometry}

\usepackage{tabularx,lipsum,environ}
\usepackage{here}
\usepackage{color}
\usepackage{hyperref}

\makeatletter
\newcommand{\problemtitle}[1]{\gdef\@problemtitle{#1}}% Store problem title
\newcommand{\probleminput}[1]{\gdef\@probleminput{#1}}% Store problem input
\newcommand{\problemquestion}[1]{\gdef\@problemquestion{#1}}% Store problem question
\NewEnviron{problem}{
  \problemtitle{}\probleminput{}\problemquestion{}% Default input is empty
  \BODY% Parse input
  \par\addvspace{.5\baselineskip}
  \noindent
  \begin{tabularx}{\textwidth}{@{\hspace{\parindent}} l X c}
    \multicolumn{2}{@{\hspace{\parindent}}l}{\@problemtitle} \\% Title
    \textbf{Input:} & \@probleminput \\% Input
    \textbf{Problem:} & \@problemquestion% Question
  \end{tabularx}
  \par\addvspace{.5\baselineskip}
}
\makeatother

\begin{document}

% \preprint{APS/123-QED}

\title[]{Computational power of one- and two-dimensional dual-unitary quantum circuits}% Force line breaks with \\

\author{Ryotaro Suzuki}
\email{ryotaro.suzuki.2139@gmail.com}
\affiliation{Graduate School of Engineering Science, Osaka University, 1-3 Machikaneyama, Toyonaka, Osaka 560-8531, Japan}
\affiliation{Dahlem Center for Complex Quantum Systems, Freie Universität Berlin, Berlin 14195, Germany}

\author{Kosuke Mitarai}
\email{mitarai@qc.ee.es.osaka-u.ac.jp}
\affiliation{Graduate School of Engineering Science, Osaka University, 1-3 Machikaneyama, Toyonaka, Osaka 560-8531, Japan}
\affiliation{Center for Quantum Information and Quantum Biology, Institute for Open and Transdisciplinary Research Initiatives, Osaka University, Osaka 560-8531, Japan}
\affiliation{JST, PRESTO, 4-1-8 Honcho, Kawaguchi, Saitama 332-0012, Japan}
 
\author{Keisuke Fujii}
\email{fujii@qc.ee.es.osaka-u.ac.jp}
\affiliation{Graduate School of Engineering Science, Osaka University, 1-3 Machikaneyama, Toyonaka, Osaka 560-8531, Japan}
\affiliation{Center for Quantum Information and Quantum Biology, Institute for Open and Transdisciplinary Research Initiatives, Osaka University, Osaka 560-8531, Japan}
\affiliation{Center for Emergent Matter Science, RIKEN, Wako Saitama 351-0198, Japan}

\begin{abstract}
Quantum circuits that are classically simulatable tell us when quantum computation becomes less powerful than or equivalent to classical computation. Such classically simulatable circuits are of importance because they illustrate what makes universal quantum computation different from classical computers.
In this work, we propose a novel family of classically simulatable circuits by making use of dual-unitary quantum circuits (DUQCs), which have been recently investigated as exactly solvable models of non-equilibrium physics, and we characterize their computational power.
Specifically, we investigate the computational complexity of the problem of calculating local expectation values and the sampling problem of one-dimensional DUQCs {whose initial states satisfy certain conditions}, and we generalize them to two spatial dimensions.
We reveal that a local expectation value of a DUQC is classically simulatable at an early time, which is linear in a system length.
In contrast, in a late time, they can perform universal quantum computation, and the problem becomes a $\sf{BQP}$-complete problem.
Moreover, classical simulation of sampling from a DUQC turns out to be hard.
\end{abstract}

\maketitle

%\tableofcontents

\section{\label{sec:Intro}Introduction}
Quantum computation is widely believed to be intractable by classical computers.
However, there also exist certain types of quantum circuits that can be efficiently simulated classically 
despite being able to generate highly entangled states.
Famous examples are quantum circuits which consist of Clifford gates \cite{gottesman1997stabilizer} or matchgates,
corresponding to free-fermionic dynamics \cite{valiant2002quantum, terhal2002classical, bravyi2004lagrangian, jozsa2008matchgates}.
Such classically simulatable quantum circuits are of importance
because they illustrate what makes universal quantum computation different from classical computers.
Moreover, they have practical applications such as randomized benchmarking \cite{magesan2011scalable, PhysRevA.85.042311, helsen2020matchgate}, simulation by stabilizer sampling \cite{howard2017application}, and estimation of an error threshold of a quantum error correction code \cite{suzuki2017efficient}. 

Classically simulatable or exactly solvable quantum circuits are also important in the study of dynamics of isolated quantum systems \cite{RevModPhys.83.863, eisert2015quantum, d2016quantum}. 
{For example, Clifford circuits include quantum dynamics which are both integrable and non-integrable
\footnote{Throughout this paper, we say that quantum dynamics are integrable if there exist an extensive number of conserved quantities. },
and they have been investigated from the perspective of quantum thermodynamics, and especially thermalization \cite{chandran2015semiclassical, gopalakrishnan2018facilitated, berenstein2021exotic}.
Moreover, physical quantities, such as entanglement entropy and out-of-time-ordered correlators, of an ensemble of one and higher dimensional Haar random unitary circuits are calculated exactly \cite{nahum2017quantum, nahum2018operator, von2018operator}.
Because, in general, it is notoriously difficult to treat quantum dynamics analytically except for one-dimensional integrable systems  \cite{essler2016quench}, above quantum circuit representations of quantum dynamics are powerful tools to analyze their physical properties. }
{In addition, since} two-dimensional quantum dynamics are not explored in comparison with one-dimensional cases, solvable models in two dimensions are in great demand.

Recently, a new class of quantum gates called {``}dual-unitary gates'' has been introduced \cite{gopalakrishnan2019unitary, PhysRevLett.123.210601}.
Dual-unitary gates are unitary gates which remain unitarity under reshuffling their indices.
The dynamics consisting of dual-unitary gates can be either integrable or non-integrable.
In Refs. \cite{PhysRevLett.123.210601, 10.21468/SciPostPhys.8.4.067}, it has been shown that dynamical correlation functions and time evolution of operator entanglement entropy under dual-unitary quantum circuits (DUQCs) are calculated exactly.
Because these quantum circuits have one spatial dimension, {we call them}  one-dimensional (1D) DUQCs. 
{Moreover}, it has been shown that when the system size is infinite, time evolution of local observables, correlation functions, and entanglement entropy of 1D DUQCs {arising from certain initial states} can be calculated exactly \cite{PhysRevB.101.094304}. 
{As we will define later, initial states are described by matrix product states (MPSs) whose matrices satisfy certain conditions. They are called solvable initial states \cite{PhysRevB.101.094304}. The simplest example is a chain of EPR pairs, and the simplest counter-example is a product state.}

Interestingly, despite the above property, dual-unitary gates contain arbitrary single-qubit gates and a certain class of two-qubit entangling gates.
Note that these gates form a universal gate set if we can apply them freely \cite{bremner2002practical}.
Nevertheless, the carefully constructed initial states and the infinite size limit allow us to compute the expectation values efficiently.
Here, we ask whether they are classically simulatable or allows universal quantum computation if the system size is finite.
{The finiteness of the system size enables us to consider the circuit depth which scales with the system size and characterize their computational power.} 

In this paper, we investigate quantum computational power of 1D and two-dimensional (2D) DUQCs.
Specifically, we characterize the computational complexity of the problem of calculating expectation values of local observables and the sampling problem of 1D and 2D DUQCs with finite system sizes.
Additionally, we study classical simulatability of correlation functions of 2D DUQCs.
{Here, 2D DUQCs takes certain initial states which are product states of solvable initial states.}
Note that the generalization to the 2D lattice is of interest not only from the viewpoint of quantum computing but also from the viewpoint that exactly solvable quantum dynamics in two spatial dimensions is limited.

\begin{figure*} [t]
    \centering
    \includegraphics[width=10cm]{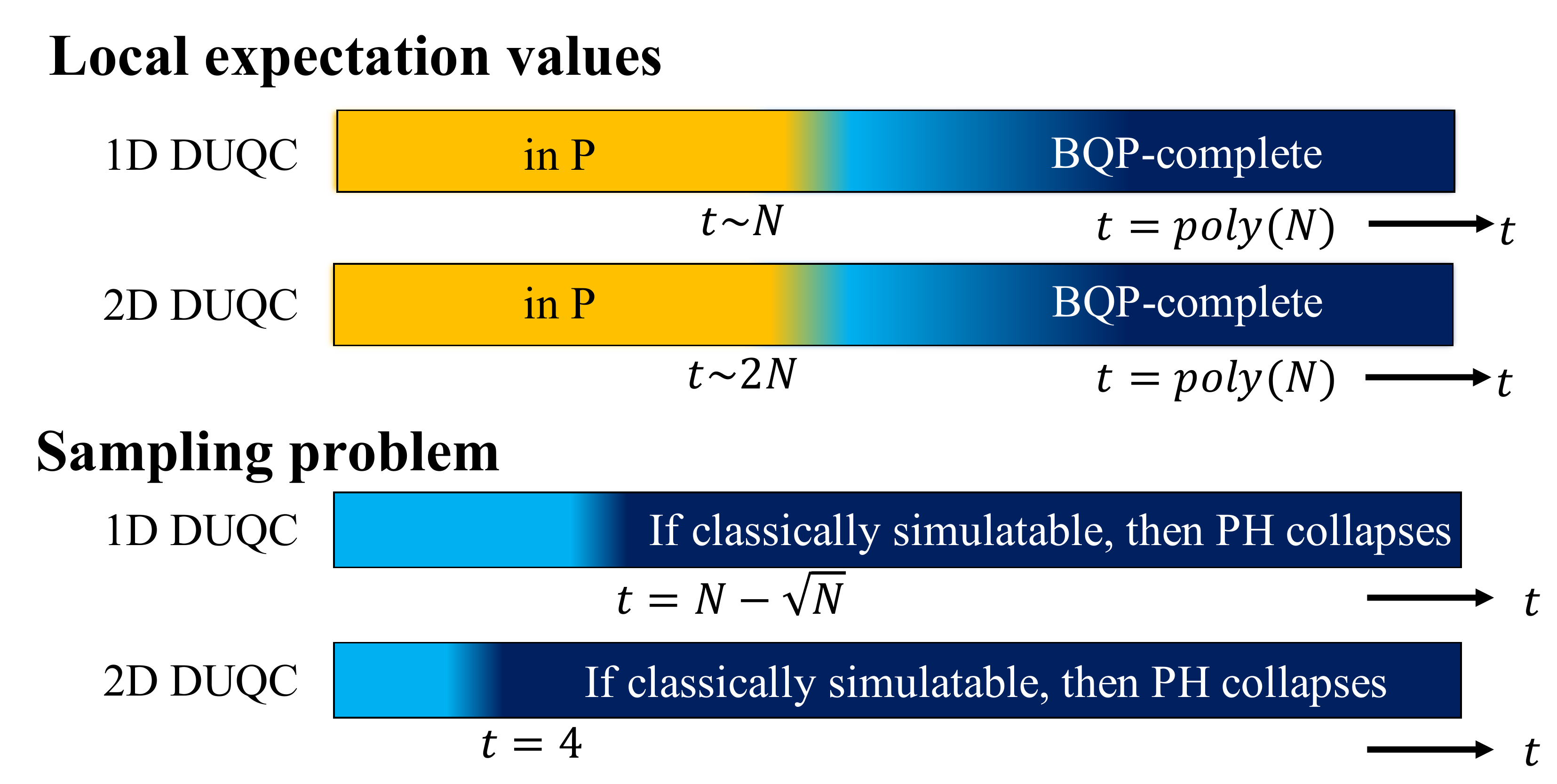}
    \caption{Summary of our main results. Computational complexity of each problem depending on the circuit depth in one and two spatial dimensions is shown.}
    \label{fig:summary}
\end{figure*}

Our results are summarized in Fig. \ref{fig:summary}.
{
For the first problem, we show that expectation values of local operators $O$ of
1D and 2D DUQCs are exponentially close to Tr($O$) until time, or circuit depth, $t \sim \frac{1}{2}N$ and $t \sim N$, respectively, where $N$ is a system length.
%In other words, local reduced density matrices of states generated by DUQCs are close to maximally mixed states.
In other words, local expectation values do not depend on specific choices of dual-unitary gates in those time regions, and they are classically simulatable.
%it means that they are classically simulatable.
On the other hand, in later time, local expectation values of DUQCs can depend on dual-unitary gates, and we find that DUQCs can simulate universal quantum computation after time poly($N$). 
It means that the problem becomes $\sf{BQP}$-complete after time poly($N$).}
{As we will discuss later, the depth overhead for simulating quantum circuits consisting of nearest-neighbor CZ gates and single-qubit gates with DUQCs is $O(N)$.}
This contrasts to conventional classically simulatable quantum circuits with a fixed gate set, such as Clifford or matchgate circuits, where  {classical simulatability does not change depending on the circuit depth.}

In addition, we show that sampling of the output of 1D and 2D DUQCs is intractable for classical computers after time $t \sim \frac{1}{2}(N-\sqrt{N})$ and $t\geq 4$, respectively, unless polynomial hierarchy (PH) collapses to its third level.
This result is based on the fact that if a quantum circuit with post-selection can simulate universal quantum computation, the output probability distribution cannot be sampled by a classical computer efficiently unless PH collapses to its third level \cite{bremner2011classical}.
It implies that, especially in the case of two dimensions, the sampling problem of constant-depth DUQCs is as hard as that of general constant-depth quantum circuits.

{Moreover, we find that correlation functions of 2D DUQCs along a special direction, {which is determined by the initial state and defined later}, become the trace of operators in linear depth, 
and hence they are classically simulatable.
In contrast, the value of correlation functions along the other direction can depend on a choice of unitary gates, and they do not seem to be classically simulatable.
We leave as an open problem whether or not the problem of calculating correlation functions of 2D DUQCs in linear depth is BQP-hard.
Finally, we also show a sufficient condition of 2D lattices on which local observables of 2D DUQCs {at an early time} are classically simulatable.
For instance, the lattices satisfying this condition include honeycomb lattices.
In summary,} we reveal that the computational power of DUQCs strongly depend on their circuit depth {and} problem settings. 
This provides a novel quantum computational model to investigate both classical simulatability of quantum computation and physical properties of non-equilibrium quantum systems.

The rest of the paper is organized as follows. 
In Sec. \ref{sec:1D}, we introduce and review 1D DUQCs. 
In Sec. \ref{sec:1D_computaional_power}, we characterize the complexity of the problem {calculating} local expectation values and the sampling problem of 1D DUQCs.
In Sec. \ref{sec:2D}, we generalize 1D DUQCs to two spatial dimensions, and characterize the complexity as with the 1D case.
After that, we discuss classical simulatability of correlation functions of 2D DUQCs and a generalization of lattices of qubits.
{Sec. \ref{sec:Conclusion} is devoted to conclusion and discussion.}

\section{\label{sec:1D} One-dimensional DUQCs}
 In this section, we review 1D DUQCs and solvable initial states, which have been introduced in Refs. \cite{gopalakrishnan2019unitary, PhysRevLett.123.210601, PhysRevB.101.094304}, and have been studied in Refs. \cite{10.21468/SciPostPhys.8.4.067,
 PhysRevX.10.031066, 
PhysRevB.102.174307, 
PhysRevResearch.2.033032, 
PhysRevB.102.064305, 
PhysRevX.11.011022, 
PhysRevLett.126.100603,
bertini2020random,
PhysRevLett.125.070501,
PhysRevResearch.3.043034,
PhysRevE.103.062133,
PhysRevB.104.014301,
hamazaki2021exceptional}.
 
\subsection{Dual-unitary gates}
 We consider a $2N$-qubit system. 
 Its computational basis is denoted by $\ket{i_1i_2\dots i_{2N}}$, where  $i_j=0,1$ indicates {a state of the $j$-th qubit}. A dual-unitary gate is a two-qubit gate in the following form:
 \begin{equation} \label{dual-unitary}
     U=e^{i\phi}u_1\otimes u_2 {\rm{SWAP}}{\rm{(CZ)}}^{\alpha}v_1 \otimes v_2,
 \end{equation}
 where SWAP is the swap gate,
  \begin{equation}{\rm{SWAP}}=
\begin{pmatrix}
1 & 0 & 0 &0\\
0 & 0 & 1 &0\\
0 & 1 & 0 &0\\
0 & 0 & 0 &1\\
\end{pmatrix},
 \end{equation}
 CZ is the controlled-Z gate,
   \begin{equation}{\rm{CZ}}=
\begin{pmatrix}
1 & 0 & 0 &0\\
0 & 1 & 0 &0\\
0 & 0 & 1 &0\\
0 & 0 & 0 &-1\\
\end{pmatrix},
 \end{equation}
 $u_1$, $u_2$, $v_1$ and $v_2$ are arbitrary single-qubit gates, and both $\phi$ and $\alpha$ are arbitrary real numbers. Alternatively, Eq. (\ref{dual-unitary}) is rewritten as
 \begin{equation}
 e^{\phi'}u_1\otimes u_2 e^{-i \frac{\pi}{4}\left( X\otimes X + Y\otimes Y +  J Z\otimes Z \right)} v_1' \otimes v_2',
  \end{equation}
 with $\phi'=\phi-\frac{\pi}{4}\alpha$, $J=\alpha +1$,  $v_1'=e^{i\frac{\pi}{4}\alpha Z}v_1$, and $v_2'=e^{i\frac{\pi}{4}\alpha Z}v_2$. It has been shown that these gates can describe both integrable and non-integrable periodically driven quantum systems, or Floquet systems \cite{PhysRevLett.123.210601}. For example, the time evolution operator of a  periodically driven quantum system with XXZ interaction,
 \begin{equation}
 e^{-i \frac{\pi}{4} \left( X\otimes X + Y\otimes Y + J Z\otimes Z \right)},
  \end{equation}
{is one of the dual-unitary gates.
The time evolution operator of a self-dual kicked Ising chain can also be written in terms of dual-unitary gates as follows:} 
  \begin{equation}
  \begin{split}
   \mathcal{T}&e^{-i \int_0^1 \left( \frac{\pi}{4}Z\otimes Z + h Z\otimes I + \frac{\pi}{4} \delta(t-1) X\otimes I \right) dt} \\
 =&e^{-i\frac{\pi}{4}}e^{-ihZ}e^{i\frac{\pi}{4} X}\otimes e^{i\frac{\pi}{4} X} \cdot e^{-i\frac{\pi}{4} Y}\otimes e^{-i\frac{\pi}{4} Y} 
 \\ &\cdot e^{-i \left( \frac{\pi}{4}X\otimes X + \frac{\pi}{4}Y\otimes Y \right)}\cdot e^{i\frac{\pi}{4} Z}\otimes e^{i\frac{\pi}{4} Z} \cdot e^{i\frac{\pi}{4} Y}e^{-ihZ} \otimes e^{i\frac{\pi}{4} Y},
   \end{split}
  \end{equation}
where $\mathcal{T}$ is the time ordered product, $\delta(t)$ is the Dirac delta function, and $h$ is a real number.
  
Dual-unitary gates have the following nice property.
Let $U$ be a nearest-neighbor two-qubit gate. 
Define $\tilde{U}$, called the dual gate of $U$, such that
 \begin{equation}
    \bra{k}\braket{l|\tilde{U}|i}\ket{j}= \bra{j}\braket{l|U|i}\ket{k}.
 \end{equation}
Then, $\tilde{U}$ is a unitary gate if and only if $U$ is a dual-unitary gate \cite{PhysRevLett.123.210601}. 
This property can be expressed graphically by using a tensor-network representation of quantum circuits. We represent a two-qubit gate $U$ and $U^{\dagger}$ as
\begin{equation}
\includegraphics[clip,width=2.0cm,valign=c]{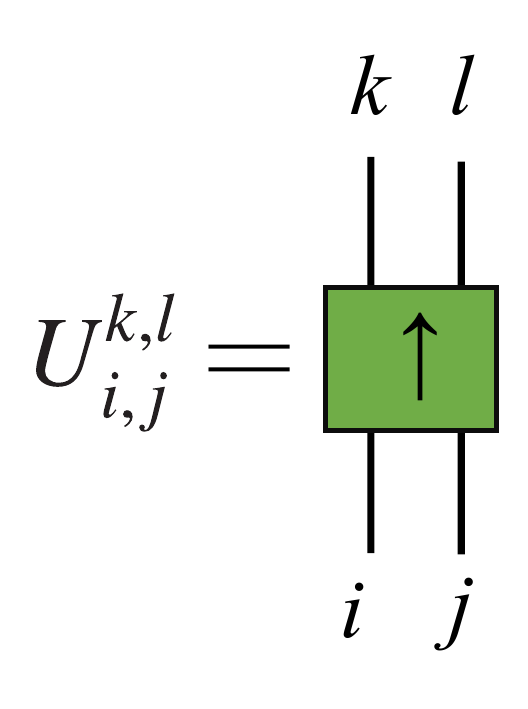},  \   \includegraphics[clip,width=2.20cm,valign=c]{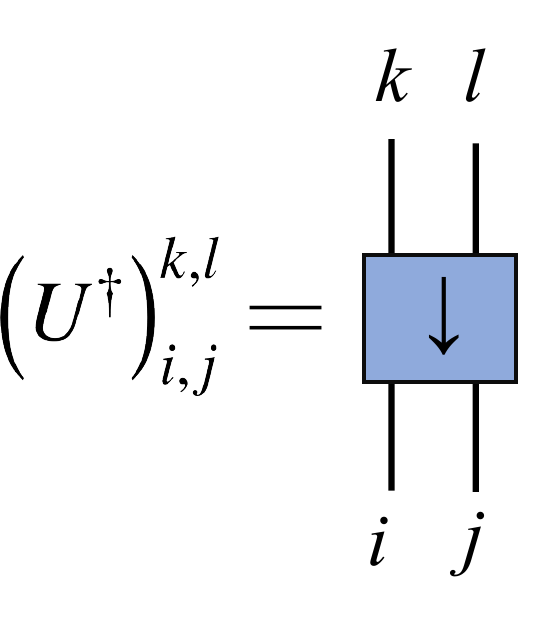},
\end{equation}
where $(i,j)$-legs and $(k,l)$-legs of $U$ and $U^{\dagger}$ serve as inputs and outputs, respectively.
The unitarity, $UU^{\dagger}=U^{\dagger}U=I$, can be written as
\begin{equation} \label{ucont}
\includegraphics[clip,width=2.8cm,valign=c]{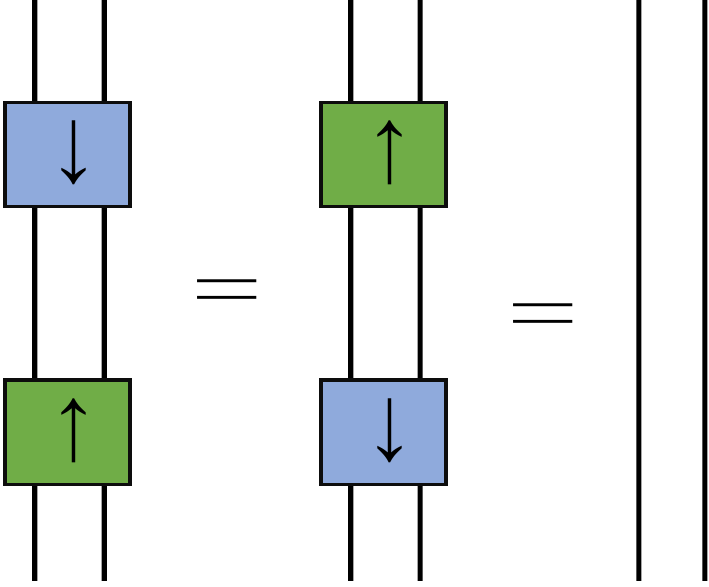}.
\end{equation}
Similarly, the dual gate of $U$ and its Hermitian conjugate are represented as
\begin{equation}
\includegraphics[clip,width=3.85cm,valign=c]{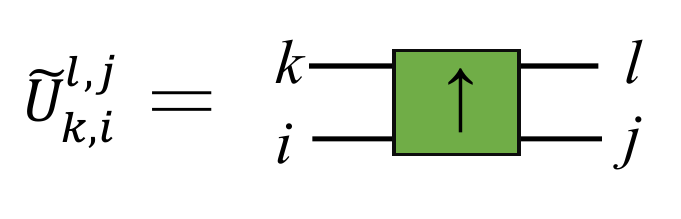},  \   \includegraphics[clip,width=4.00cm,valign=c]{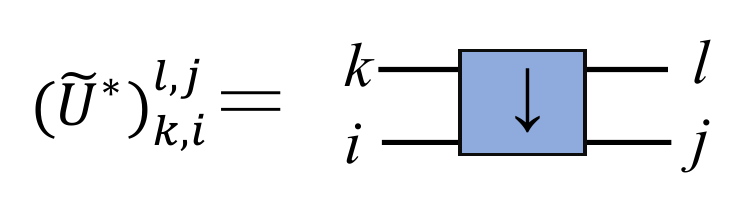},
\end{equation}
where $(i,j)$-legs and $(k,l)$-legs of $\tilde{U}$ and $\tilde{U^{\dagger}}$ serve as input and outputs, respectively.
The property of a dual-unitary gate, namely $\tilde{U}\tilde{U}^{\dagger}=\tilde{U}^{\dagger}\tilde{U}=I$, can be written as 
\begin{equation} \label{ducont}
\includegraphics[clip,width=3.4cm,valign=c]{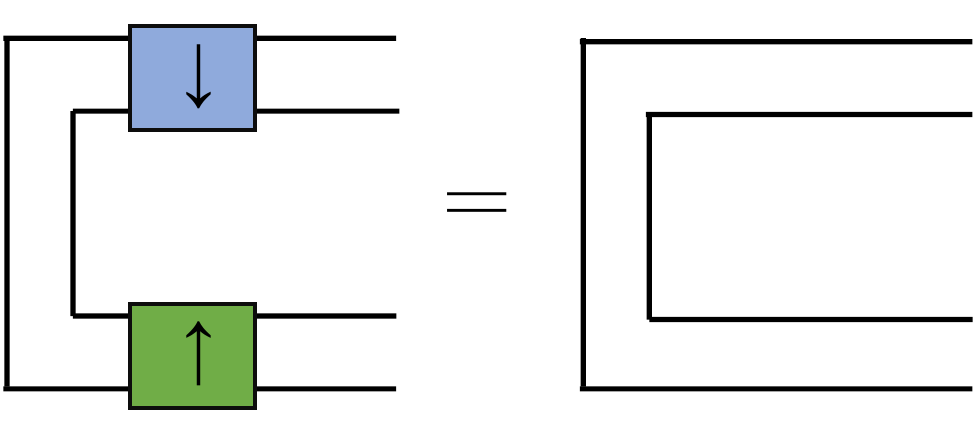}, \ \ \ \ \ 
\includegraphics[clip,width=3.4cm,valign=c]{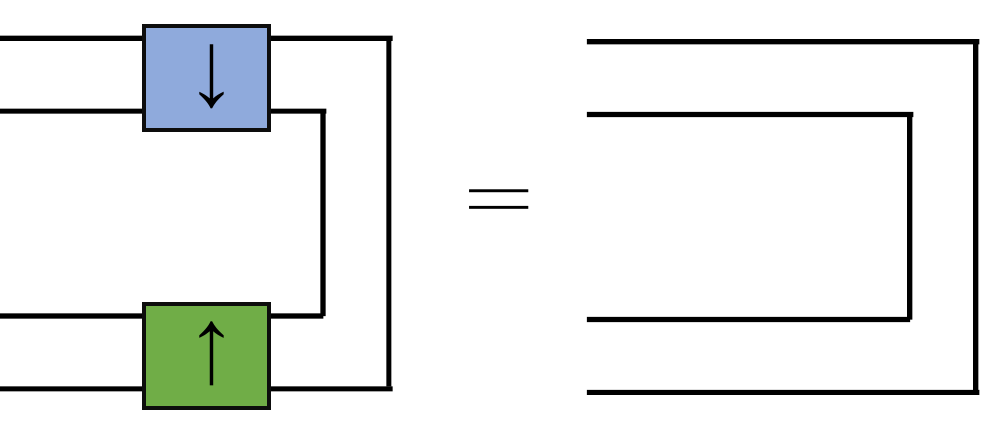}.
\end{equation}

\subsection{Dual-unitary quantum circuits}
1D DUQCs are quantum circuits with $2N$ qubits which consist of nearest-neighbor dual-unitary gates.
They are defined as follows:
\begin{equation} 
\bm{U_{\rm{1D}}}(t)=\prod_{\tau=1}^{t/2}U^{(e)}(2\tau)U^{(o)}(2\tau-1), \label{eq.1Dqc}
\end{equation}
where
\begin{align}
& U^{(o)}(2\tau-1)=\prod_{i=1}^{N}U_{2i,2i+1}(2\tau-1), \label{eq.1Dqc/o} \\
& U^{(e)}(2\tau)=\prod_{i=1}^{N}U_{2i-1,2i}(2\tau),\label{eq.1Dqc/e}
\end{align}
$t$ is an even number, and $U_{i,j}(\tau)$ is a dual-unitary gate acting on {qubit $i$ and $j$ } at time $\tau$,
or graphically,
\begin{equation}
\bm{U_{\rm{1D}}}(t) = 
\includegraphics[clip,width=6cm,valign=c]{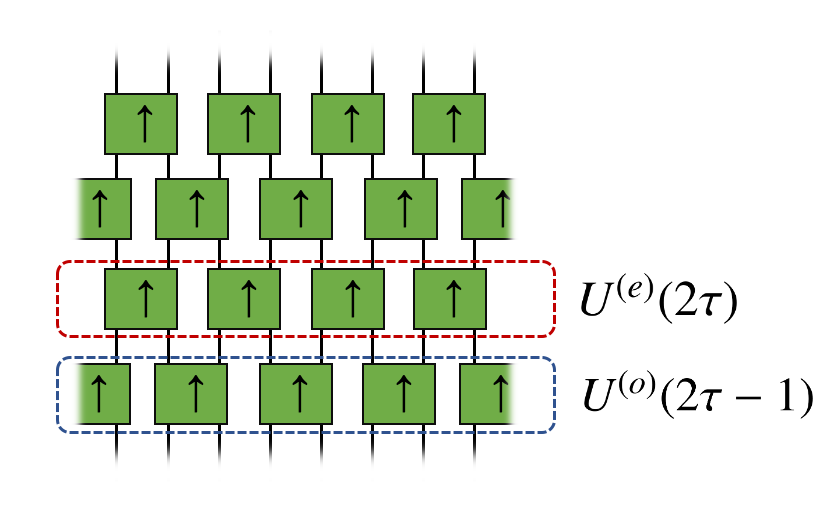}. \label{fig.1Dqc}
\end{equation}
We note that dual-unitary gates can differ from each other, that is, {the} quantum dynamics can be inhomogeneous in space and time. In Eqs. (\ref{eq.1Dqc}) {to} (\ref{fig.1Dqc}), we assume a periodic boundary condition {(PBC)} in space, that is, $U_{2N, 2N+1}=U_{2N, 1}$. 

\subsection{Solvable initial states}
In a 1D DUQC, when an initial state satisfies certain conditions and a system size is infinite, it has been shown that time evolution of local observables, correlation functions and entanglement entropy can be calculated exactly \cite{PhysRevB.101.094304}.
Such an initial state is called a solvable initial state and can be described in terms of a matrix product state (MPS) \cite{10.5555/2011832.2011833, PhysRevB.101.094304}.

% An MPS becomes solvable initial states for dual unitary circuits if its transfer matrix has a unique maximum eigenvalue and whose matrix can be regarded as a unitary gate (see Eq. (\ref{matrix_unitarity})) \cite{PhysRevB.101.094304}.
Here we describe two conditions that make a two-site shift invariant MPS,
\begin{equation}
\ket{\Psi^N(A)}=\sum_{\{i_j\}} \mathrm{Tr} \left( A^{(i_1,i_2)}A^{(i_3,i_4)}\dots A^{(i_{2N-1},i_{2N})}\right) \ket{i_1i_2\dots i_{2N}},
\end{equation}
where $A^{(i,j)}$ is a $\chi$-dimensional square matrix, { a solvable initial state.}
$\ket{\Psi^N(A)}$ can alternatively be represented by a tensor-network as,
\begin{align}
&\includegraphics[clip,width=3.8cm,valign=c]{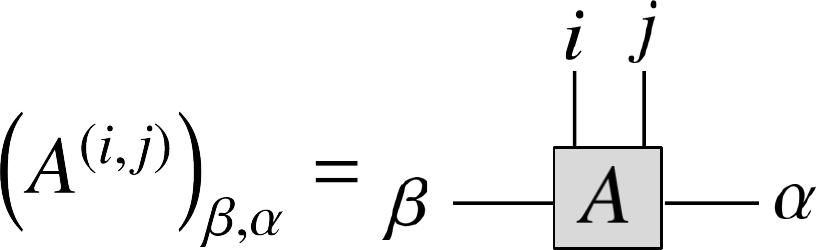}, \\
&\includegraphics[clip,width=5.8cm,valign=c]{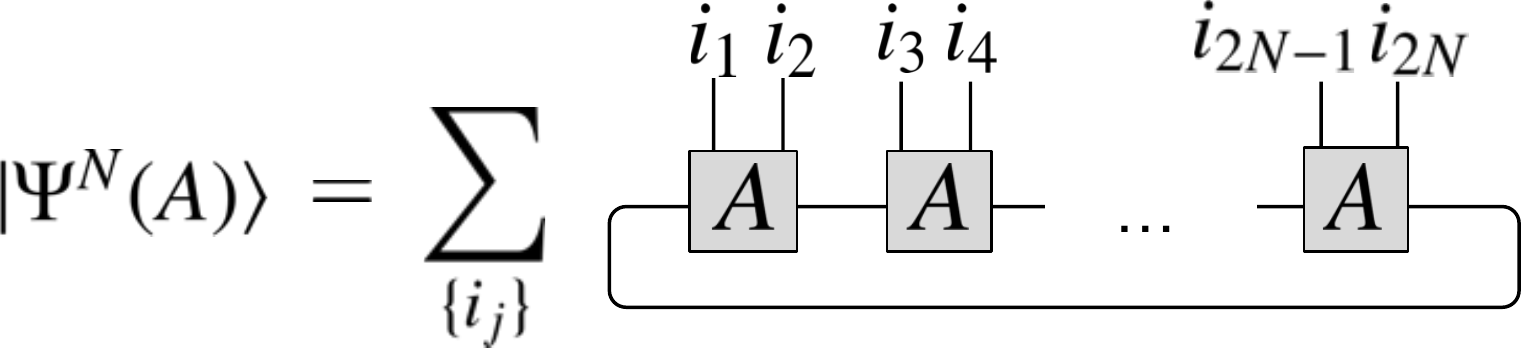}.
\end{align}
The first condition is that its transfer matrix has a unique eigenvector with a maximum eigenvalue $\lambda_0$. 
A transfer matrix associated with $\ket{\Psi^N(A)}$ is defined as
\begin{align}
E_{\beta'\beta,\alpha'\alpha}=\sum_{i,j}(A^{(i,j)*})_{\beta',\alpha'} \otimes (A^{(i,j)})_{\beta,\alpha},
\end{align}
or graphically,
\begin{align}\label{transfer}
\includegraphics[clip,width=3.8cm, valign=c]{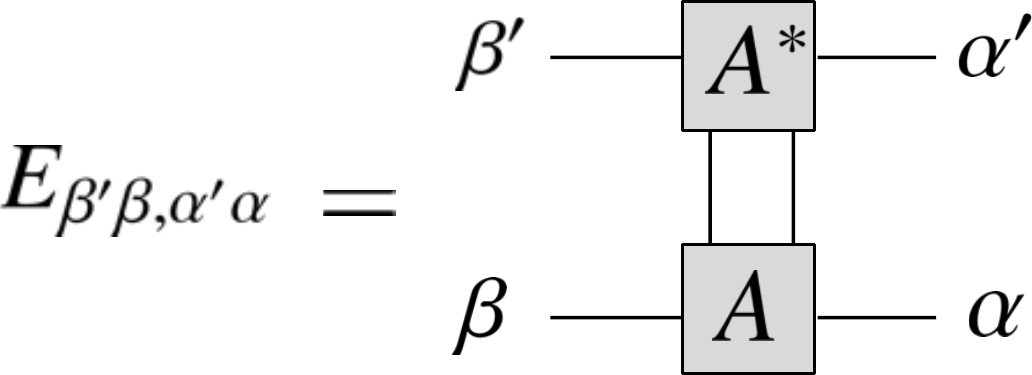}.
\end{align}
Note that $\braket{\Psi^N|\Psi^N}={\rm{Tr}}(E^N) \approx \lambda_0^N$ for large $N$, which implies $\lambda_0=1$ is needed in order to normalize $\ket{\Psi^N(A)}$, namely $\braket{\Psi^N|\Psi^N}=1$ in the limit of $N\to\infty$.
The second condition is that $A$ satisfies the following condition:
\begin{equation} \label{condition}
\sum_{k=1}^2 A^{(i,k)}(A^{(j,k)})^{\dagger}=\frac{1}{2}\delta_{i,j},
\end{equation}
or graphically,
\begin{align} \label{matrix_unitarity}
\includegraphics[clip,width=3.8cm,valign=c]{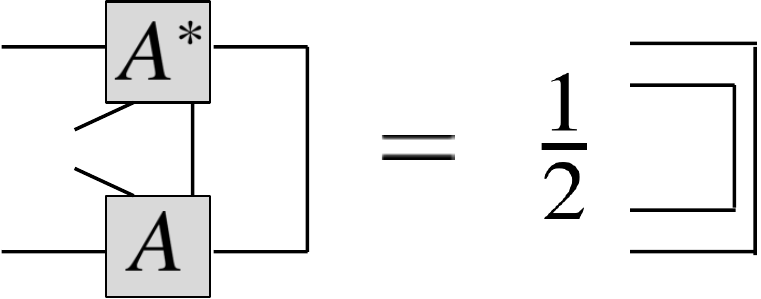}, 
\end{align}
where $\delta_{i,j}$ is Kronecker's delta. 
Eq. (\ref{condition}) implies 
\begin{equation} \label{condition'}
\sum_{k=1}^2 (A^{(i,k)})^{\dagger}A^{(j,k)}=\frac{1}{2}\delta_{i,j}.
\end{equation}
Furthermore, Eqs. (\ref{condition}) and (\ref{condition'}) imply that the transfer matrix has
\begin{equation}
\ket{I}=\frac{1}{\sqrt{\chi}}\sum_{\alpha=1}^\chi\ket{\alpha \alpha},
\end{equation}
as right and left eigenvectors with an eigenvalue 1.
Strictly speaking, the second condition considered here is more restrictive than that of Ref. \cite{PhysRevB.101.094304}.
However, they are equivalent in the thermodynamic limit (see Theorem 1 in Ref. \cite{PhysRevB.101.094304}), and we adopt Eq. (\ref{matrix_unitarity}) for clarity.

The simplest example of solvable states is a chain of EPR pairs ${\ket{{\rm{EPR}}}}^{\otimes N}$, where
$\ket{{\rm{EPR}}}=\frac{1}{\sqrt{2}}(\ket{00}+\ket{11})$,
or graphically,
\begin{equation} \label{EPR}
\includegraphics[clip,width=7cm,valign=c]{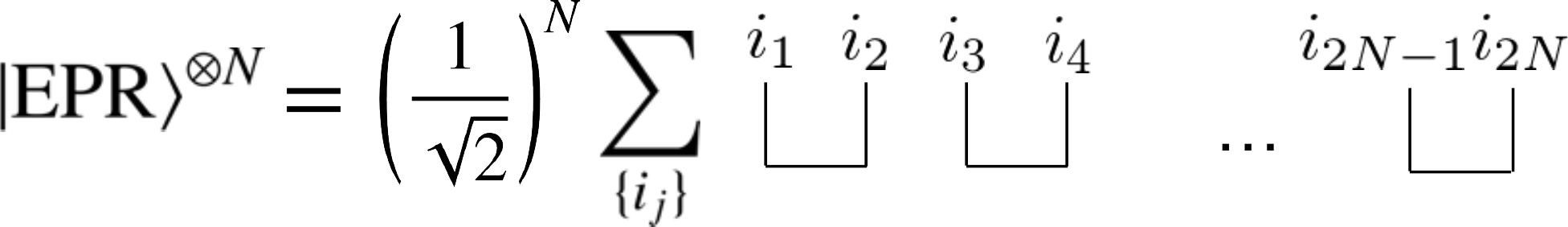},
\end{equation}
whose $\chi$ is zero.
This example has been studied to analyze the entanglment dynamics of self-dual kicked Ising chains \cite{bertini2019entanglement}.

\subsection{{Local expectation values of 1D DUQCs}}
Let us briefly describe how expectation values of local observables can be calculated for dual unitary circuits with solvable initial states \cite{PhysRevB.101.094304}.
We consider a time-evolved transfer matrix $E(t)$ defined as
\begin{equation} \label{E(t)}
\includegraphics[clip,width=4cm,valign=c]{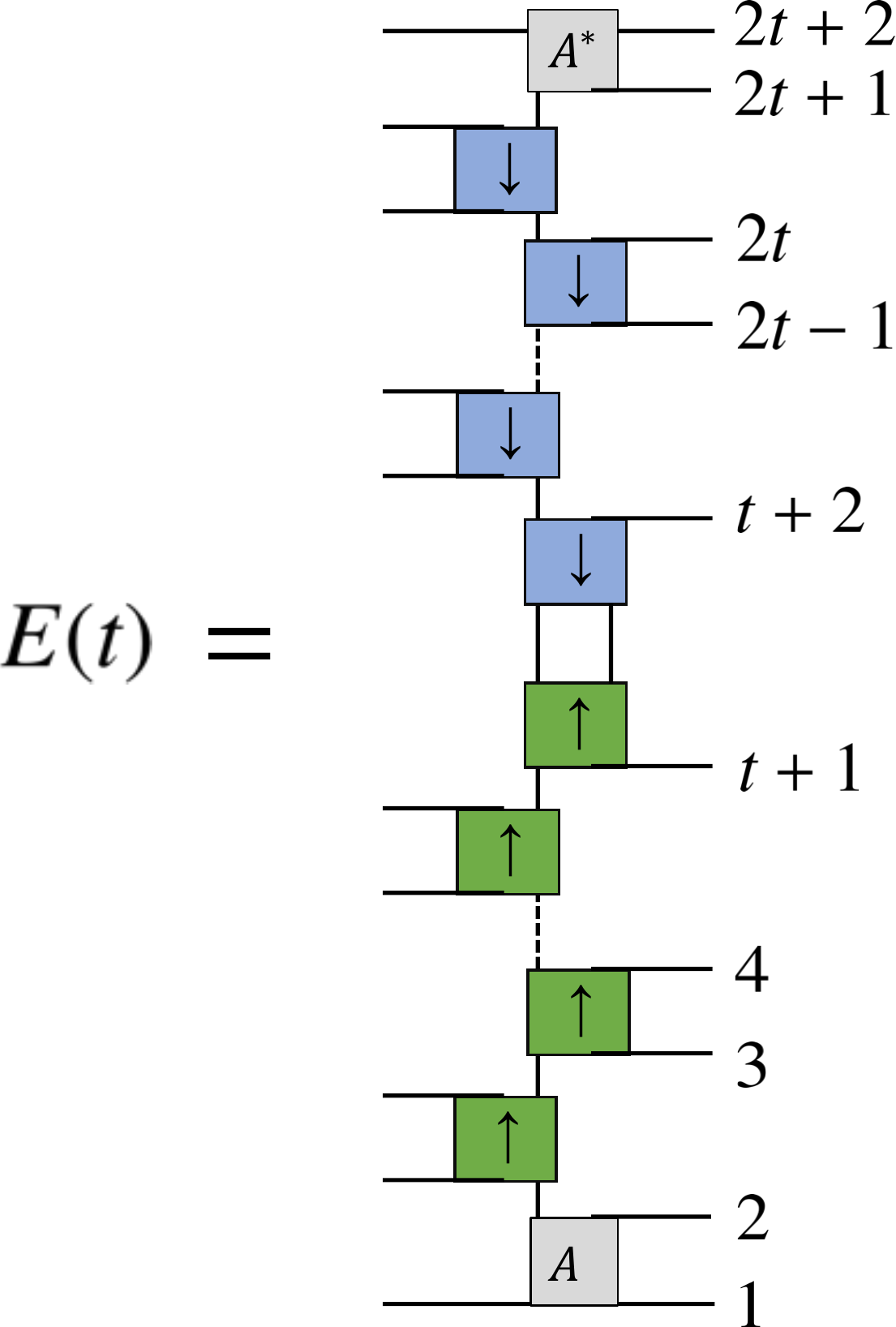},
\end{equation}
where each number of the right-side $2t+2$ legs indicates the input space which $E(t)$ {acts} on.
It can be shown by Eqs. (\ref{ducont}) and (\ref{matrix_unitarity}) that
\begin{equation}\label{eq:E(t)-eigenvector}
\ket{I(t)}=\bigotimes_{j=2}^{t+1} \left(\frac{1}{\sqrt{d}}\sum_{i=1}^2\ket{i}_j\ket{i}_{2t-j+3}  \right) \otimes \left(\frac{1}{\sqrt{\chi}}\sum_{\alpha=1}^\chi\ket{\alpha}_1\ket{\alpha}_{2t+2}\right)
\end{equation}
is both right and left eigenvectors of $E(t)$ with eigenvalue 1. 
In fact, $\ket{I(t)}$ is the unique eigenvector with maximum eigenvalue.
This leads to the following equality:
  \begin{equation} \label{shortcorr}
\lim_{N \to \infty} \left( E(t)^N\right)=\ket{I(t)}\bra{I(t)}.
  \end{equation}
By virtue of Eq. (\ref{shortcorr}), one can calculate an expectation value of a local observable $O$ as follows:
\begin{align}
\includegraphics[clip,height=5cm, valign=c]{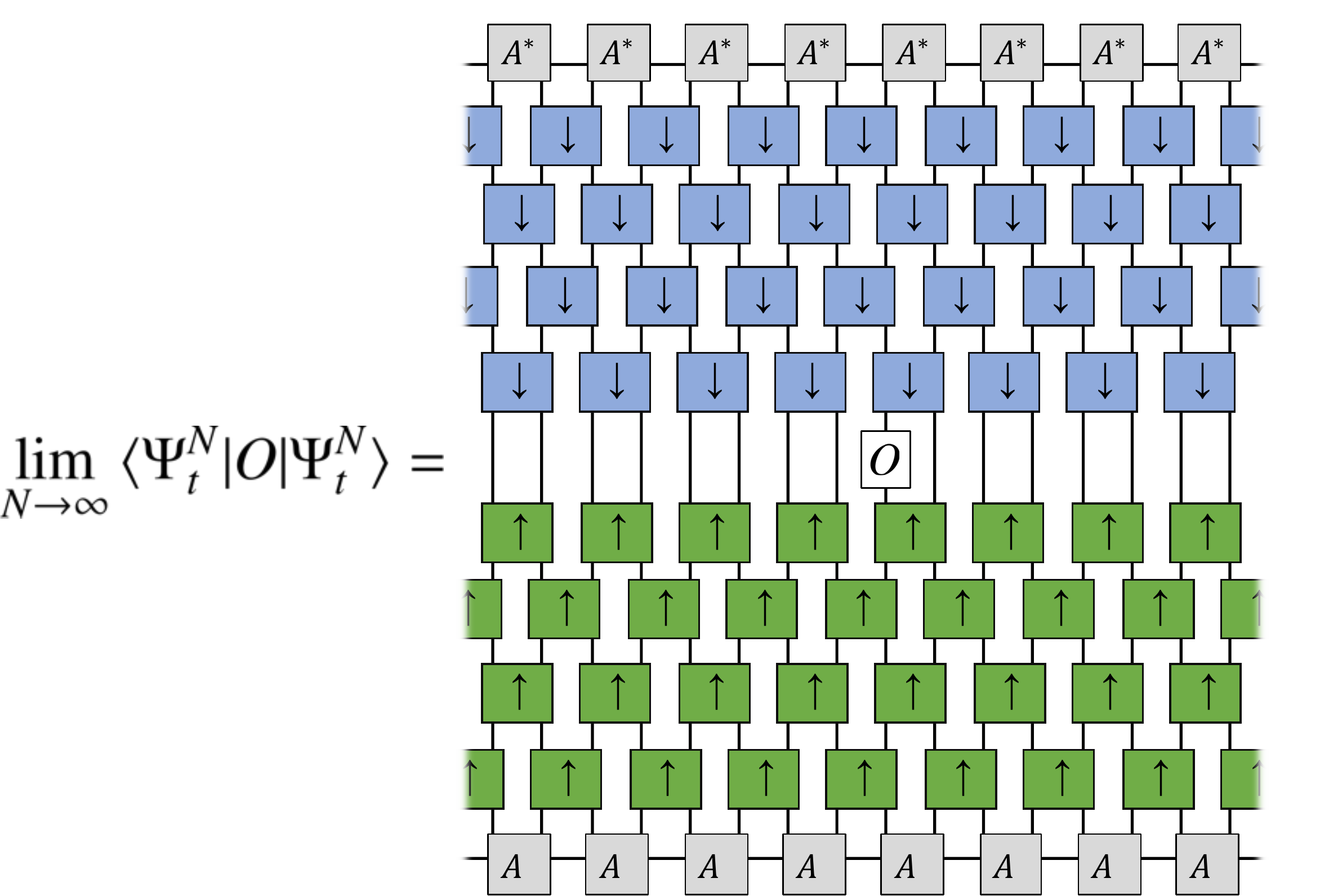} \nonumber \\
\includegraphics[clip,height=5cm,valign=c]{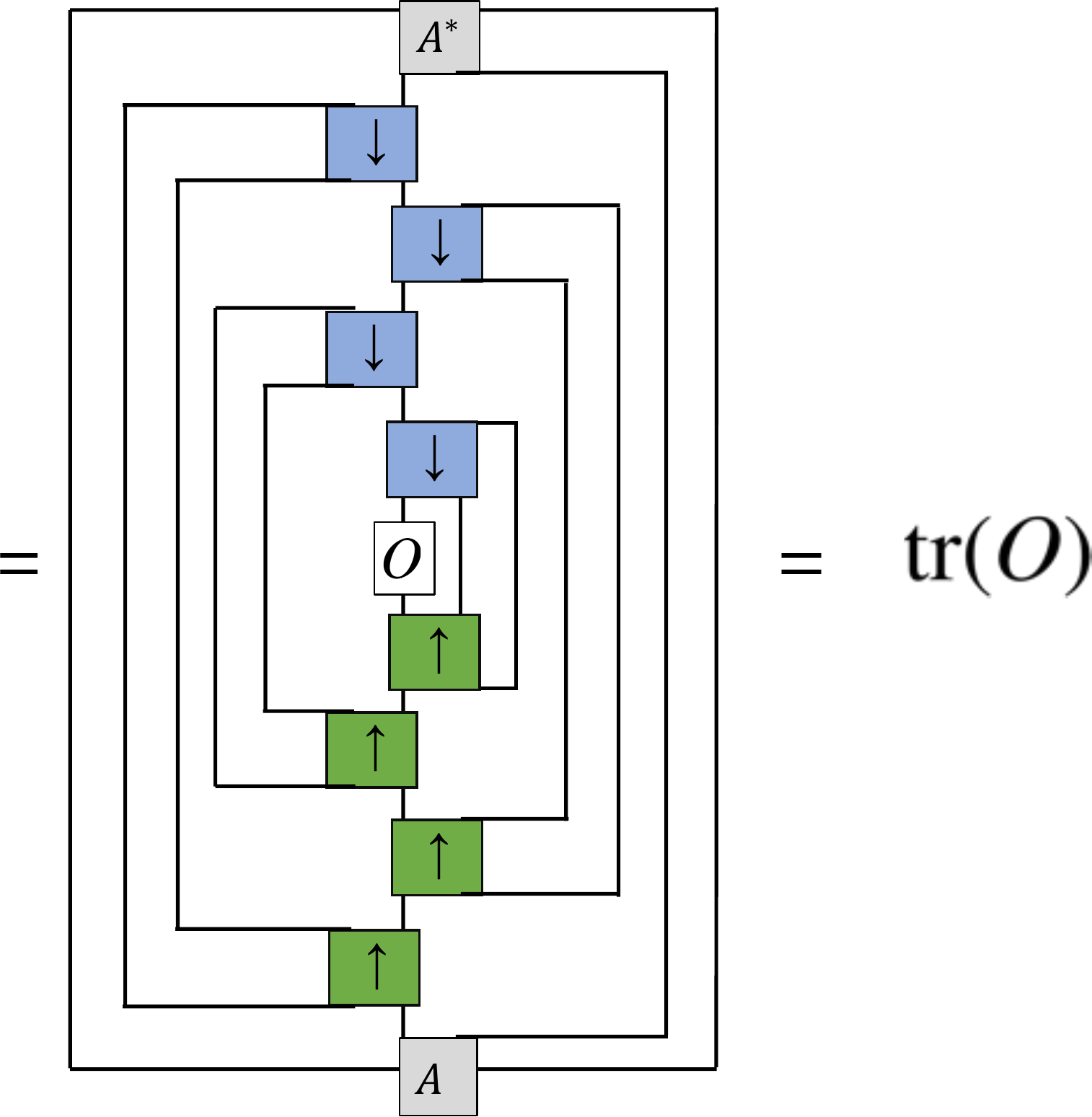},
\end{align}
where $\ket{\Psi^N_t}=\bm{U_{\rm{1D}}}(t)\ket{\Psi^N(A)}$ is a solvable initial state evolved to time $t$.
It means that an expectation value of a local observable does not depend on a specific choice of dual-unitary gates.
{From a similar argument,} it has been also shown that correlation functions of 1D DUQCs are classically simulatable \cite{PhysRevB.101.094304}.

\section{\label{sec:1D_computaional_power} computational power of 1D DUQC}
\subsection{\label{subsec:1D_obs} Local expectation values}
Although local observables in a 1D DUQC are calculated exactly when the system size is infinite, a dual-unitary gate can contain arbitrary single-qubit gates and the CZ gate, which can form a universal gate set in principle \cite{PhysRevA.52.3457}.
Thus, it is natural to ask whether or not DUQCs can perform universal quantum computation when their system size is finite. 
{Finiteness of the system size enables us to consider the circuit depth which is scaling with the system size and characterize their computational power.} 

In this section, we answer the above question affirmatively and characterize computational complexity of the problem of {calculating} local expectation values for dual-unitary circuits.
More precisely, we consider a local observable with length $l$
\begin{equation}
O=\prod_{i=0}^{l-1} O_{i_0+i},
\end{equation}
where $i_0$ is {an} integer, and $O_{i_0+i}$ is an observable on qubit $i_0+i$. 
Local expectation values of $O$ at time $t$ $\braket{O(t)}$ is $\braket{O(t)}=\frac{\braket{\Psi^N_t|O|{\Psi^N_t}}}{\braket{\Psi^N_t|\Psi^N_t}}$. 
Here, we note that $\ket{\Psi^N(A)}$ is not generally normalized for a finite $N$.
Then, we define the following decision problem, which has a parameter $t$.

\begin{dfn}[local expectation values of 1D DUQCs] \label{1DLE}
Consider a 1D DUQC in time $t$ $\bm{U_{\rm{1D}}}(t)$, {a} solvable $2N$-qubit initial state $\ket{\Psi^N(A)}$ with $\chi = O(1)$, and {a} local observable $O$ whose operator norm is 1 with length $l = O(1)$.
$\braket{O(t)}$ is promised to be either $\geq a$ or  $\leq b$, where $a-b \geq \frac{1}{poly(N)}$. 
The problem is to determine whether $\braket{\Psi^N_t|O|{\Psi^N_t}}$ is $\geq a$ or  $\leq b$.
\end{dfn}

%\begin{problem}
%  \problemtitle{\textsc{local expectation values of 1D DUQCs {at an early time}}}
%  \probleminput{a DUQC $\bm{U_{\rm{1D}}}(t)$ with $t\leq N+1-l/2$, 
%  \newline solvable $2N-$qubit initial state $\ket{\Psi^N(A)}$,
%  \newline local observable $O$ whose operator norm is 1
%  \newline with length $l = O(1)$.}
%  \problemquestion{$\braket{\Psi^N_t|O|{\Psi^N_t}}$ is promised to be either $\geq \frac{2}{3}$ or  $\leq \frac{1}{3}$. 
%  \newline Determine whether $\braket{\Psi^N_t|O|{\Psi^N_t}}$ is $\geq \frac{2}{3}$ .}
%\end{problem}

%\begin{problem}
%  \problemtitle{\textsc{local observables of 1D DUQCs in poly time}}
%  \probleminput{a DUQC $\bm{U_{\rm{1D}}}(t)$ with $t=\rm{poly}(N)$, 
%  \newline solvable $2N-$qubit initial state $\ket{\Psi^N(A)}$,
%  \newline local observable $O$ whose operator norm is 1 
%  \newline with length $l = O(1)$.}
%  \problemquestion{$\braket{\Psi^N_t|O|{\Psi^N_t}}$ is promised to be either $\geq \frac{2}{3}$ or $\leq \frac{1}{3}$.  
%  \newline Determine whether  $\braket{\Psi^N_t|O|{\Psi^N_t}}$ is $\geq \frac{2}{3}$ .}
%\end{problem}

 We show that {\bf{Problem} \ref{1DLE}} with time $t \leq \lfloor (1-\delta)N \rfloor - {l/2}$, for {an arbitrary} $0< \delta <1$,  is in $\sf{P}$.
{In contrast, with time $t={\rm{poly}}(N)$, it  becomes $\sf{BQP}$-complete.}
We first prove the above statements in the case where the initial state is a chain of EPR pairs Eq. (\ref{EPR}), which is normalized for any $N$, for simplicity.
Then, we extend to general solvable initial states.

\subsubsection{{1D DUQC is classically simulatable at an early time}}
First, we prove the former statement.
In this case, an expectation value $\braket{\Psi^N_t|O|\Psi^N_t}$ with time $t \leq N-{l/2}$ can be written as 
\begin{equation} \label{Evalue1}
\includegraphics[clip,width=7.5cm,valign=c]{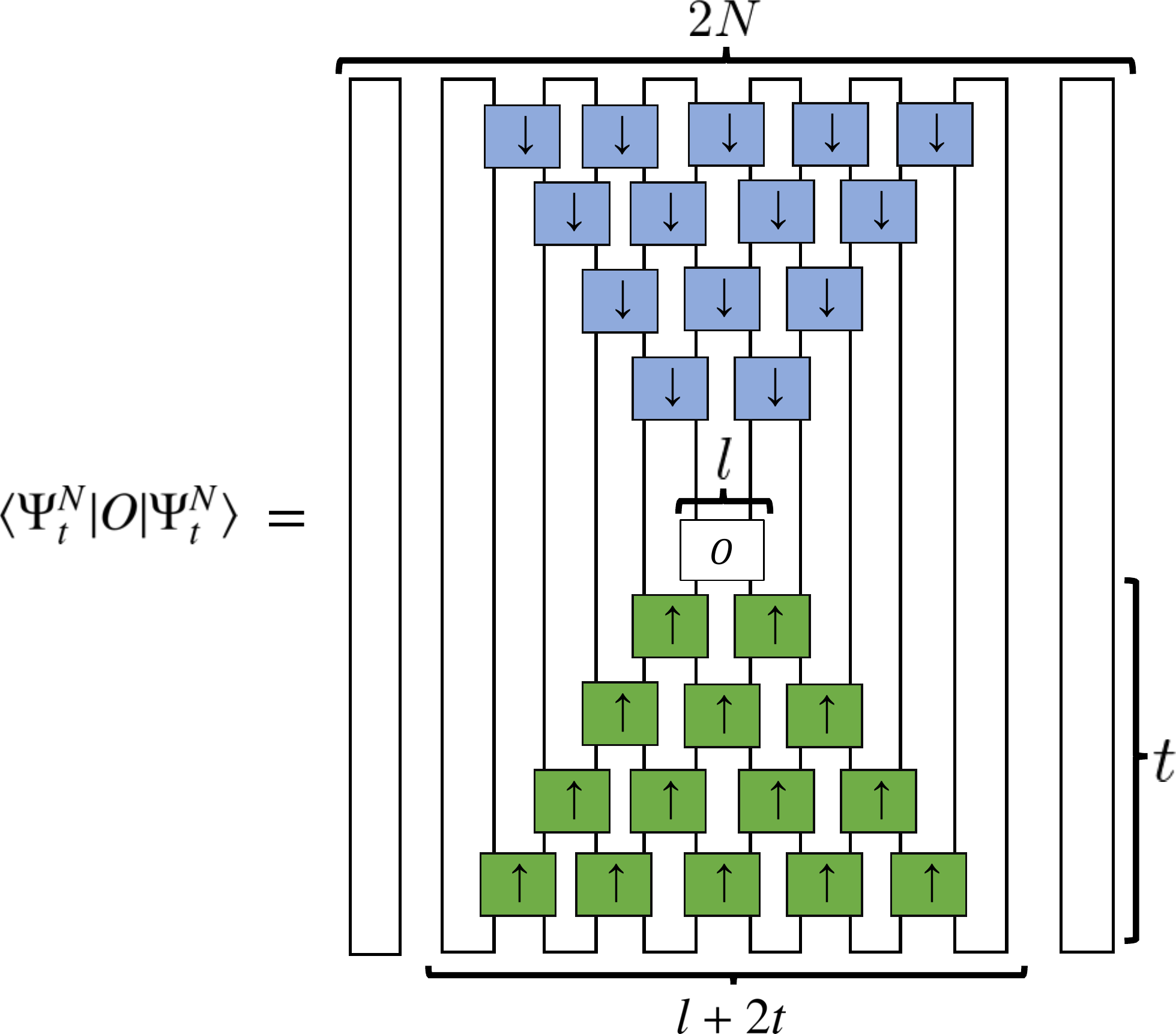},
\end{equation}
where $l$ and $i_0$ are respectively assumed to be even for clarity, and the normalization coefficient which arises from Eq. (\ref{EPR}) is omitted.
To derive Eq. (\ref{Evalue1}), we remove unitary gates outside of the causal-cone by using Eq. (\ref{ucont}).
Although we fixed parity of $l$ and $i_1$, expectation values with other combinations of parity can be written likewise.
If $t \geq \frac{1}{2}l+1$, we obtain $\braket{\Psi^N_t|O|\Psi^N_t}={\rm{Tr}}(O)$ by adapting Eq. (\ref{ducont}) to the both sides of Eq. (\ref{Evalue1}) sequentially and removing leftover unitary gates by adapting Eq. (\ref{ducont}).
In other words, an expectation value $\braket{O(t)}$ is identical to the expectation value of a maximally mixed state regardless of components of dynamics.
The detail of the proof is in Appendix \ref{sec:appA}.
{For $t<l/2 + 1$, in general, the local expectation value is not equal to Tr$(O)$ as the dual-unitary gates cannot be cancelled.}
However, local expectation values {at time  less than}  $l/2+1$ are classically simulatable because only constant number of unitary gates are involved in calculation of local expectation values. 

We can easily generalize the above results to general solvable initial states. Let us note that a transfer matrix of a solvable MPS to the power $M$ can be written as 
\begin{equation} \label{finite}
E^M=\ket{I}\bra{I}+{\varepsilon^M},
\end{equation}
where {$\varepsilon^M$} is a matrix such that leading order of non-zero elements are $O(|\lambda_1|^M)$ and $\lambda_1$ is the second largest eigenvalue of $E$. By using Eq. (\ref{finite}), an expectation value of $O$ with $t \leq \lfloor (1-\delta)N \rfloor - {l/2}$ can be calculated as follows:
\begin{equation} \label{Evalue3}
\includegraphics[clip,width=9cm]{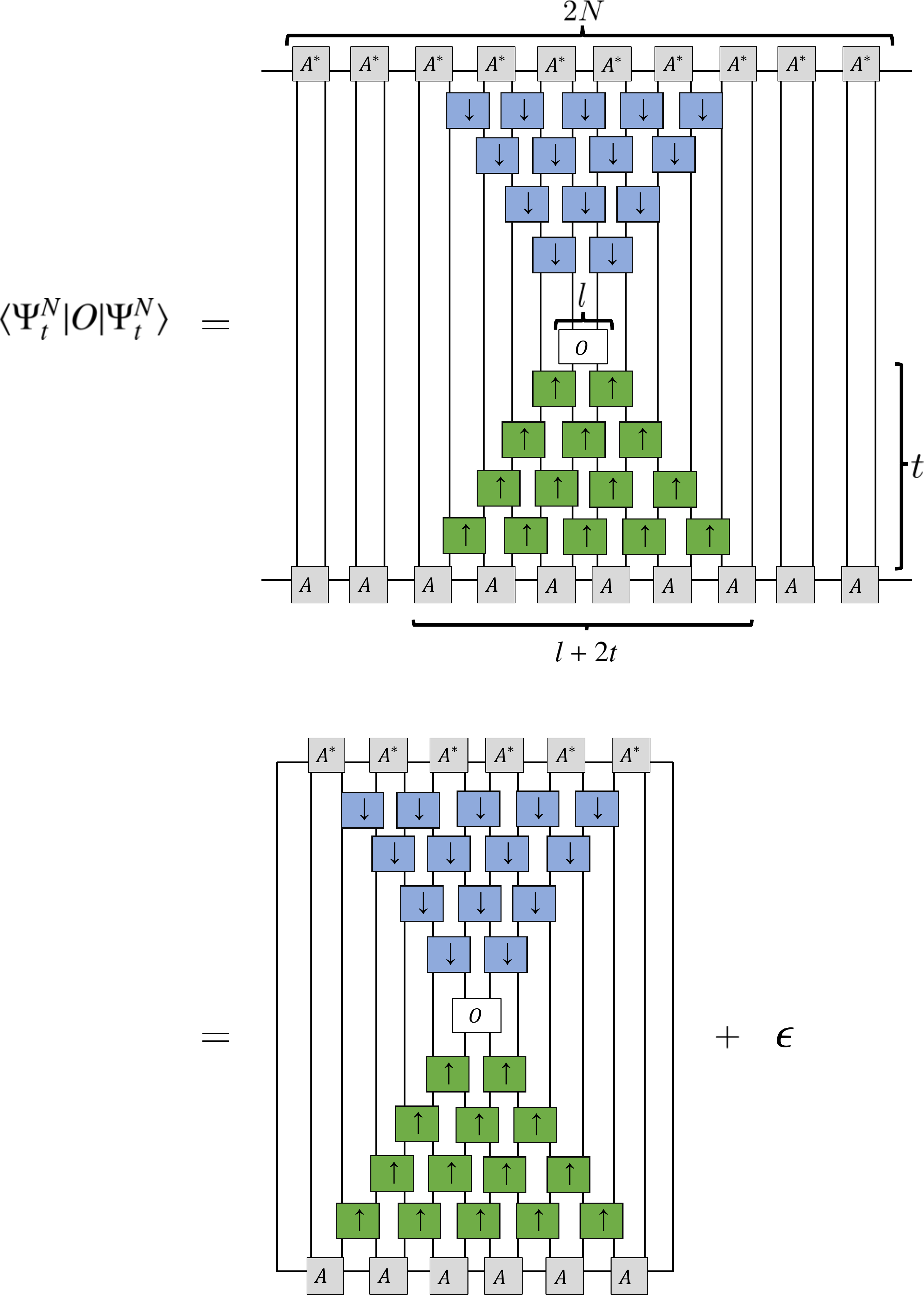},
\end{equation}
where $\epsilon$, deriving from the matrix {$\varepsilon^M$}, 
is $O \left( |\lambda_1|^{\lfloor \delta N \rfloor} \right)$.
Then, by applying Eqs. (\ref{ucont}), (\ref{ducont}), (\ref{condition}), and (\ref{condition'}) to Eq. (\ref{Evalue3}), we obtain the expectation value $\braket{O(t)}={\rm{Tr}}(O)+\epsilon$.
The detail of the calculation is in Appendix \ref{sec:appB}.
Because $\epsilon$ is exponentially suppressed with respect to $N$, {\bf{Problem \ref{1DLE}}} with $t \leq \lfloor (1-\delta)N \rfloor - {l/2}$ is still in $\sf{P}$. 
Here, we note that if {an initial state} is a chain of EPR pairs, $\delta$ can be chosen as zero, that is, $\braket{O(t)}$ with $t \leq N-{l/2}$ is classically simulatable.

On the other hand, for time $t > N-{l/2}$, the expectation value can be written as
\begin{equation} \label{Evalue2}
\includegraphics[clip,width=4.5cm,valign=c]{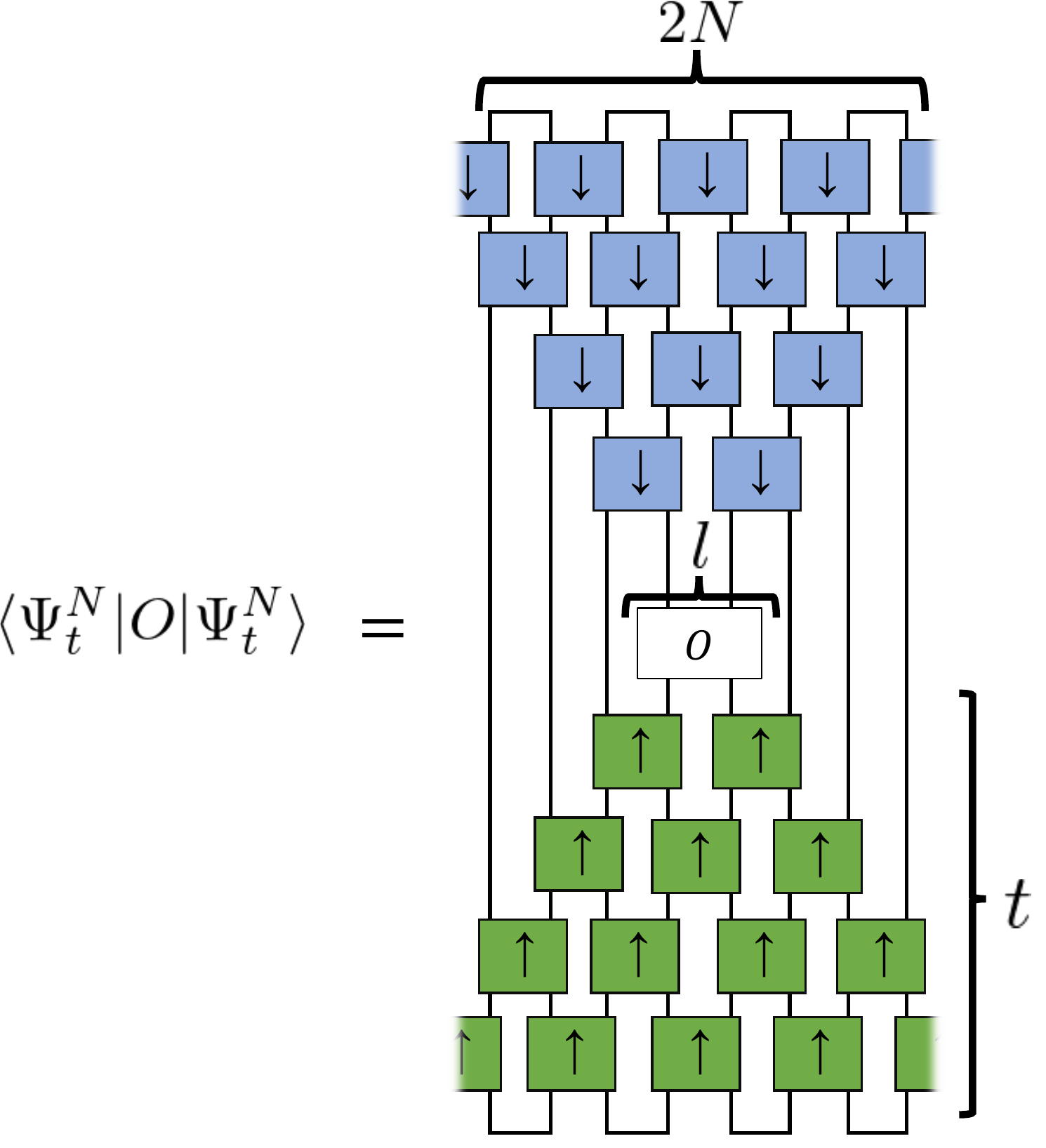}.
\end{equation}
In contrast to Eq. (\ref{Evalue1}), we cannot use Eqs. (\ref{ucont}) and (\ref{ducont}) to calculate Eq. (\ref{Evalue2}) efficiently.
It could be possible that we can calculate Eq. (\ref{Evalue2}) efficiently by another method.
 In the following section, we exclude the possibility by showing {\bf{Problem} \ref{1DLE}} with $t={\rm{poly}}(N)$ is BQP-complete.

\subsubsection{{1D DUQC is universal in late time}}
The inclusion of the problem in $\sf{BQP}$ is trivial; we can just execute the dual-unitary circuit to obtain an expectation value.
What is left to be shown is that the problem is $\sf{BQP}$-hard. 
To do that, we consider a $\sf{BQP}$-complete problem {calculating} whether an local expectation value $c_n=\braket{0^n|U^{\dagger}\frac{I+Z_1}{2}U|0^n}$ is $\geq a $, where $c_n$ is promised to be either $\geq a$ or $\leq b$ , where $a-b \geq \frac{1}{poly(N)}$, and $U$ is a 1D quantum circuit consisting of poly($n$) nearest-neighbor two-qubit gates.
Then, we show that {\bf{Problem} \ref{1DLE}} is as hard as the $\sf{BQP}$-complete problem after time poly($N$).
 
 \begin{figure} [t]
    \centering
\includegraphics[clip,width=8.5cm]{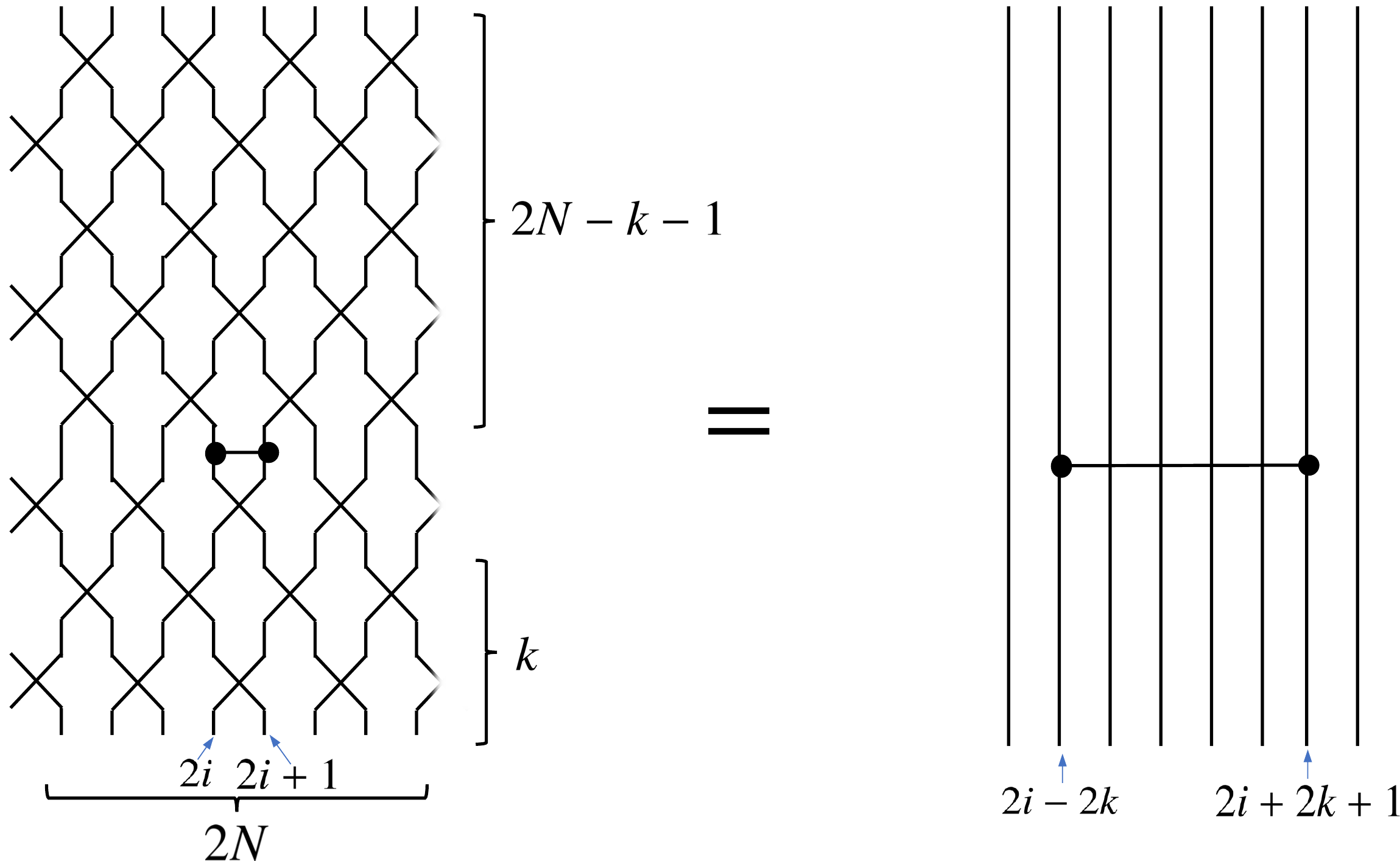}.
    \caption{Graphical representation of Eq. (\ref{CZl1}). The indices in this figure correspond to those in Eq. (\ref{CZl1}).}
    \label{fig:swap_cz}
\end{figure}

 Firstly, we construct the CZ gate acting on an even numbered site and an odd numbered site by cancelling swap gates from a DUQC, as follows:
\begin{align}
&{\rm{CZ}}_{2i-2k,2i+2k+1} \notag \\
=&\prod_{t_2=1}^{N-k-1} \left( \bf{SWAP_2}\bf{SWAP_1} \right) \bf{SWAP_2} \bf{CZ_{2i,2i+1}}
\cdot \prod_{t_1=1}^{k} \left( \bf{SWAP_2}\bf{SWAP_1} \right), \label{CZl1}
\end{align}
\begin{align}
&{\rm{CZ}}_{2i-2k-1,2i+2k+2} \notag \\
=&\prod_{t_2=1}^{N-k-1} \left( \bf{SWAP_2}\bf{SWAP_1} \right) \bf{CZ_{2i-1,2i}} \bf{SWAP_1}
\cdot \prod_{t_1=1}^{k} \left( \bf{SWAP_2}\bf{SWAP_1} \right), \label{CZl2}
\end{align}
where
\begin{equation}
{\bf{SWAP_1}}=\prod_{i=1}^{N}{\rm{SWAP}}_{2i,2i+1}, 
\end{equation}
\begin{equation}
{\bf{SWAP_2}}=\prod_{i=1}^{N}{\rm{SWAP}}_{2i-1,2i}, 
\end{equation}
\begin{equation} \label{CZ_even}
{\bf{CZ_{2i,2i+1}}}= \left( \prod_{j=1}^{i}{\rm{SWAP}}_{2j,2j+1} \right) {\rm{CZ}}_{2i,2i+1} \left( \prod_{j=i+1}^{N}{\rm{SWAP}}_{2j,2j+1} \right),
\end{equation}
\begin{equation} \label{CZ_odd}
{\bf{CZ_{2i-1,2i}}}= \left( \prod_{j=1}^{i}{\rm{SWAP}}_{2j-1,2j} \right) {\rm{CZ}}_{2i-1,2i} \left( \prod_{j=i+1}^{N}{\rm{SWAP}}_{2j-1,2j} \right).
\end{equation}
Remember that the index $2N+1$ is treated as $1$ according to a {PBC}.
Graphically, Eq. (\ref{CZl1}) can be written as Fig. \ref{fig:swap_cz}. 
{Additionally, one can apply CZ gates in parallel by substituting ${\rm{CZ}} \cdot {\rm{SWAP}}$ gates for some of $\rm{SWAP}$ gates in Eqs. (\ref{CZ_even}) and (\ref{CZ_odd}).}

Secondly, {it can be easily shown that an arbitrary product of single-qubit gates} can be implemented by {substituting suitable $u_1\otimes u_2 \cdot {\rm{SWAP}}$ gates for some of ${\rm{SWAP}}$ gates} in Eqs. (\ref{CZl1}) and (\ref{CZl2}). 
A quantum circuit consisting of arbitrary one-qubit gates and CZ gates has capability to efficiently simulate arbitrary quantum circuit consisting of poly($n$) nearest-neighbor two-qubit gates \cite{PhysRevA.52.3457}.
Therefore, we can construct a DUQC $\bm{U_{\rm{1D}}}$ with poly($N$) depth such that {
\begin{equation}
\bra{{\rm{EPR}}}^{\otimes N} \bm{U_{\rm{1D}}}^{\dagger} \left( \frac{I+Z_1}{2} \right) \bm{U_{\rm{1D}}} \ket{\rm{EPR}}^{\otimes N}=c_n
\end{equation}
holds}, which means that {\bf{Problem \ref{1DLE}}} with time $t =$ poly$(N)$ and input a chain of EPR  pairs is a $\sf{BQP}$-complete problem.
{Here, we remark that, from our construction, if $U$ is written as a depth-d quantum circuit consisting of nearest-neighbor CZ gates and single-qubit gates, the depth of  $\bm{U_{\rm{1D}}}$ which simulates $U$ is $O(dN)$.  
This is because $\bm{U_{\rm{1D}}}$ requires $O(N)$ simulation cost for CZ gates in $U$ in each time.
It means that the depth overhead for simulating $U$ with $\bm{U_{\rm{1D}}}$ is $O(N)$.
}

\subsection{\label{subsec:1D_samp} Sampling problem}

 We now move on to discuss the sampling complexity of 1D DUQCs. 
 We show that classical simulation of sampling from linear depth 1D DUQCs is hard. 

Let $\{ p_z \}$ be the probability distribution with ${p_z}$ being the probability of obtaining output $z \in \{ 0,1 \} ^{2N}$ when $\ket{{\Psi}_{t}}$ is measured in computational basis. 
We define that a probability distribution $\{ p_z\}$ is sampled by classical computers efficiently with a multiplicative error $c$
if there exists a classical probabilistic polynomial-time algorithm that outputs $z$ with
probability $q_z$ such that $|p_z-q_z| \leq c p_z$, for all $z$. 
It has been shown that if an $n$-qubit quantum circuit with post-selection can simulate universal quantum computation, 
the measurement output of $n$ qubits cannot be sampled by classical computers efficiently with a multiplicative error 
unless polynomial hierarchy collapses to its third level \cite{bremner2011classical}. 
On the other hand, measuring a square lattice cluster state in an arbitrary measurement basis with post-selection can perform universal quantum computation \cite{PhysRevLett.86.5188}. 
A square lattice cluster state is defined as
\begin{equation}
    \ket{\rm{CS}}=\prod_{(i,j) \in E}{\rm{CZ}}_{i,j} \ket{+}^{\otimes N},
\end{equation}
where $\ket{+}$ is $H\ket{0}$, we assigned a qubit in $\ket{+}$ state to each vertex of a square lattice, and $E$ is the set of all edges of the lattice.

Here, we prove that a 1D DUQC can generate a square lattice cluster state with an EPR-chain initial state in time $N-\sqrt{2N}/2+1$, which implies that sampling from $\bm{U_{\rm{1D}}}\ket{\mathrm{EPR}}^{\otimes N}$ is classically hard.
% A square lattice cluster state mapped to one dimension can be generated by depth-$(N-\sqrt{2N}/2+1)$ DUQCs with the initial state $\ket{\rm{EPR}}$.

\begin{figure*} [t]
    \includegraphics[width=15cm]{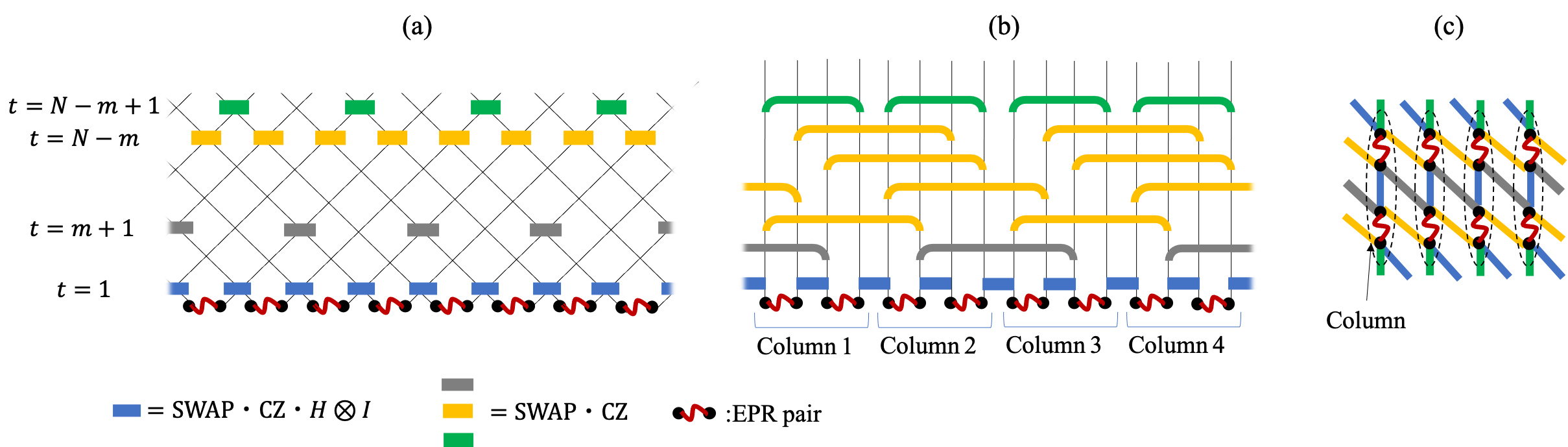}
    \caption{A DUQC which generate a square lattice cluster state mapped to one dimension. 
    The DUQC in Eq.(\ref{1dclu,1}), (\ref{1dclu,2}), (\ref{1dclu,3}), and (\ref{1dclu,4}) can be written as (a), and it is equal to a quantum circuit (b), which include quantum gates acting on two distant qubits. The final state of (b) is equivalent to a square lattice cluster state (c) by deviding a 1D qubit array into $2m$ and rearranging it to a square lattice.}
    \label{fig:1D;cluster}
\end{figure*}

We assume that $2N$ is a square of an even number $2m$, which enables us to make a one-to-one correspondence between the 1D qubits and ones on square lattice.
Note that the following equation holds:
\begin{equation}
    I\otimes H \ket{\rm{EPR}}=\rm{CZ} \ket{++},
\end{equation}
where $H$ is the Hadamard gate.
With this in mind, we obtain a square lattice cluster state by applying the following dual-unitary circuit to the initial state $\ket{\rm{EPR}}$:
\begin{align}
U_{\mathrm{cluster}}=\prod_{t=1}^{t=N-m+1}U_{\mathrm{cluster}}(t),
\end{align}
where,
\begin{align}
& U_{\mathrm{cluster}}(t=1)=\prod_{i=1}^{N}({\rm{SWAP}\cdot \rm{CZ}}\cdot H\otimes I)_{2i, 2i+1},\label{1dclu,1} \\
& U_{\mathrm{cluster}}(t=m+1)=\prod_{j=1}^{2m}\prod_{i=1}^{m-1}({\rm{SWAP}\cdot \rm{CZ}})_{2i+m+j, 2i+m+1+j} \cdot {\rm{SWAP}}_{3n+j, 3n+i+j},\label{1dclu,2} \\
& U_{\mathrm{cluster}}(t=N-m)=\prod_{i=1}^{N}({\rm{SWAP}\cdot \rm{CZ}})_{2i, 2i+1}, \label{1dclu,3}\\
& U_{\mathrm{cluster}}(t=N-m+1)=\prod_{i=1}^{2m}({\rm{SWAP}\cdot \rm{CZ}}\cdot H\otimes I)_{N+m+j, N+m+1+j}, \label{1dclu,4}
\end{align}
and $U_{\mathrm{cluster}}(t)$ at other odd and even times are ${\bf{SWAP_1}}$ and ${\bf{SWAP_2}}$, respectively. 
Graphically, the above DUQC can be written as Fig. \ref{fig:1D;cluster} (a), and it is equivalent to Fig. \ref{fig:1D;cluster} (b). 

As shown in Fig. \ref{fig:1D;cluster} (b) and (c), by rearranging the 1D qubit array (b) to the square lattice (c), the final state of the DUQC is equivalent to the square lattice cluster state.
Moreover, because dual-unitary gates include arbitrary single-qubit gates, we can measure the final state in an arbitrary measurement basis.
Therefore, sampling from depth-$(N-\sqrt{2N}/2+1)$ 1D DUQCs is unlikely to be classically simulatable.

\section{\label{sec:2D} Generalization to 2D DUQCs}
In this section, we generalize DUQCs to two spacial dimensions and characterize their computational power.

\subsection{\label{subsec_def} Definition of 2D DUQCs} 
We consider a $2N \times 2M$-qubit system on a $2N \times 2M$ square lattice. 
Its computational basis is denoted by $\ket{i_{(1,1)} i_{(1,2)} \cdots i_{(1,2M)}i_{(2,1)} \cdots  i_{(2N,2M)}}$, where $i_{(j,k)}=0,1$ indicates a state of the $(j,k)$-th qubit. 

We define $U^{(1)}$, $U^{(2)}$, $U^{(3)}$, and $U^{(4)}$ as
\begin{equation}
U^{(1)}=\prod_{j=1}^{N}\prod_{k=1}^{2M}U^{\rm{D}}_{(2j,k),(2j+1,k)}, \label{u1}
\end{equation}
\begin{equation}
U^{(2)}=\prod_{j=1}^{2N}\prod_{k=1}^{M}U_{(j,2k-1),(j,2k)},\label{u2}\\
\end{equation}
\begin{equation}
U^{(3)}=\prod_{j=1}^{N}\prod_{k=1}^{2M}U^{\rm{D}}_{(2j-1,k),(2j,k)},\label{u3}
\end{equation}
\begin{equation}
U^{(4)}=\prod_{j=1}^{2N}\prod_{k=1}^{M}U_{(j,2k),(j,2k+1)},\label{u4}
\end{equation}
where $U^{\rm{D}}_{(i,j),(k,l)}$ is a dual-unitary gate acting on qubit $(i,j)$ and qubit $(k,l)$, and $U_{(i,j),(k,l)}$ is {an} arbitrary two-qubit unitary gate acting on qubit $(i,j)$ and qubit $(k,l)$. 
We also define $U^{\{2,4\}}$ as a matrix which is an arbitrary product of $U^{(2)}$ and $U^{(4)}$, for example, $U^{(2)}$, $U^{(2)} U^{(4)}$, or $U^{(2)} U^{(4)} U^{(2)}$. 
We note that the fact that a unitary gate in $U^{(2)}$ or $U^{(4)}$ can be an arbitrary unitary gate is a significant difference between one and two spatial dimentions. 
In Eqs. (\ref{u1}) {to} (\ref{u4}), we assumed a {PBC} in space, that is, $U_{(2N,k),(2N+1,k)}=U_{(2N,k),(1,k)}$ and $U_{(j,2M),(j,2M+1)}=U_{(j,2M),(j,1)}$ for all $k$ and $j$.
Then, {we define} 2D DUQCs are quantum circuits with $2N \times 2M$ qubits as follows:
\begin{equation} \label{2D_dual_unitary}
\bm{U_{\rm{2D}}}(t)=\prod_{{\tau}=1}^{\frac{t}{4}}U^{\{2,4\}}(\tau+3)U^{(3)}(\tau+2)U^{\{2,4\}}(\tau+1)U^{(1)}(\tau),
\end{equation}
where $t$ is a multiple of four and $U^i(\tau)$, $i=1,\{2,4\}, 3$, is a unitary $U^i$ at time $\tau$. 

In the following {subsections}, we consider initial states $\ket{{\mathbf{\Psi}}_A}$ which are product states of solvable initial states $\ket{{\Psi}^N_A}$ aligned in rows:
\begin{equation}
\ket{{\mathbf{\Psi}}_A}=\ket{{\Psi}^N_A}^{\otimes 2M},
\end{equation}
or graphically,
\begin{equation} \label{2D_sol_state} 
\includegraphics[clip,width=7cm]{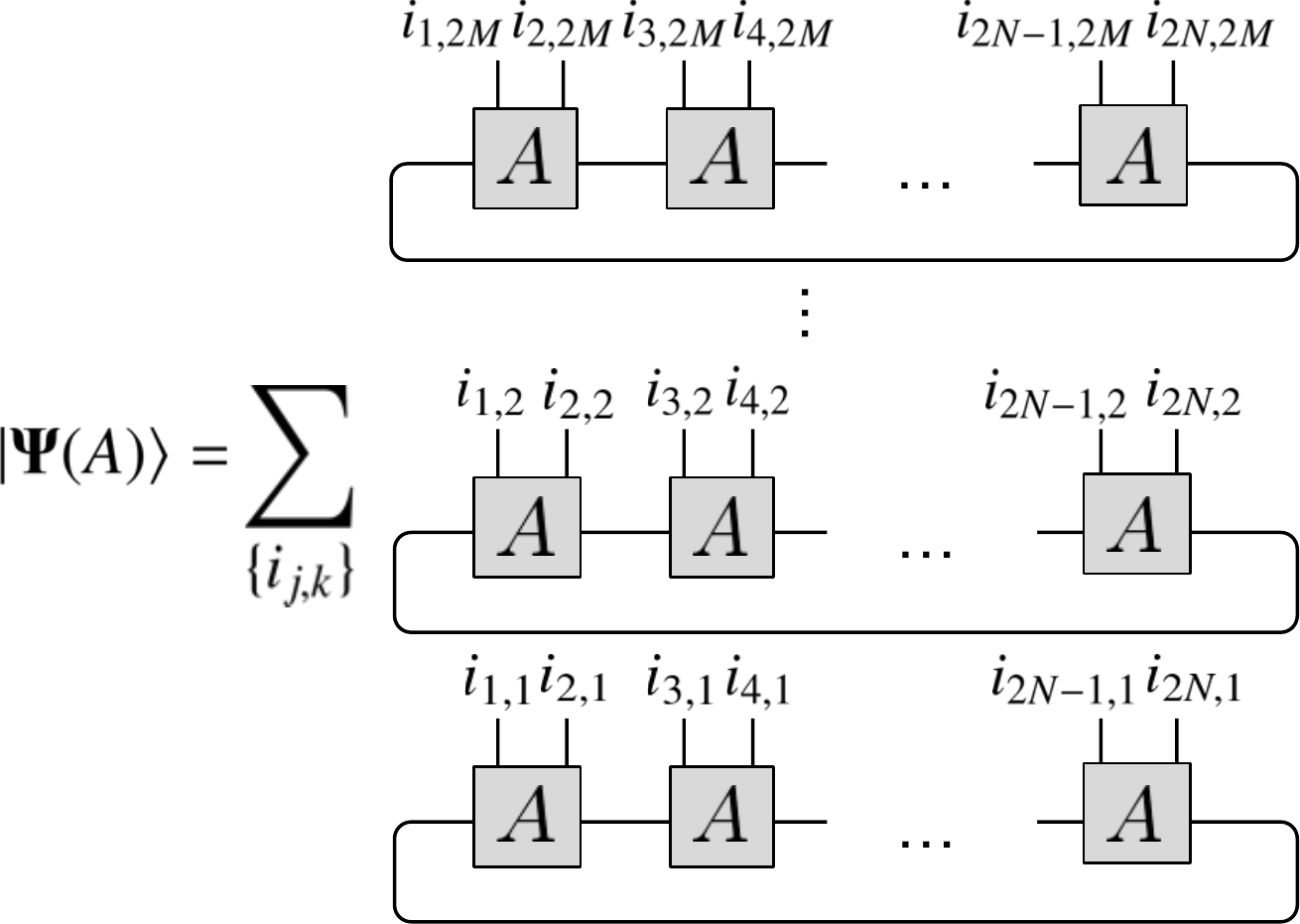},
\end{equation}
where $i_{j,k}$ indicates the state of ($j,k$)-th qubit.
We call these initial states 2D solvable initial states in the sense that, as we show in the next section, local expectation values are classically simulatable {at an early time}.

The simplest example of 2D solvable initial states is rows of chains of EPR pairs
\begin{equation}
\includegraphics[clip,width=7.5cm]{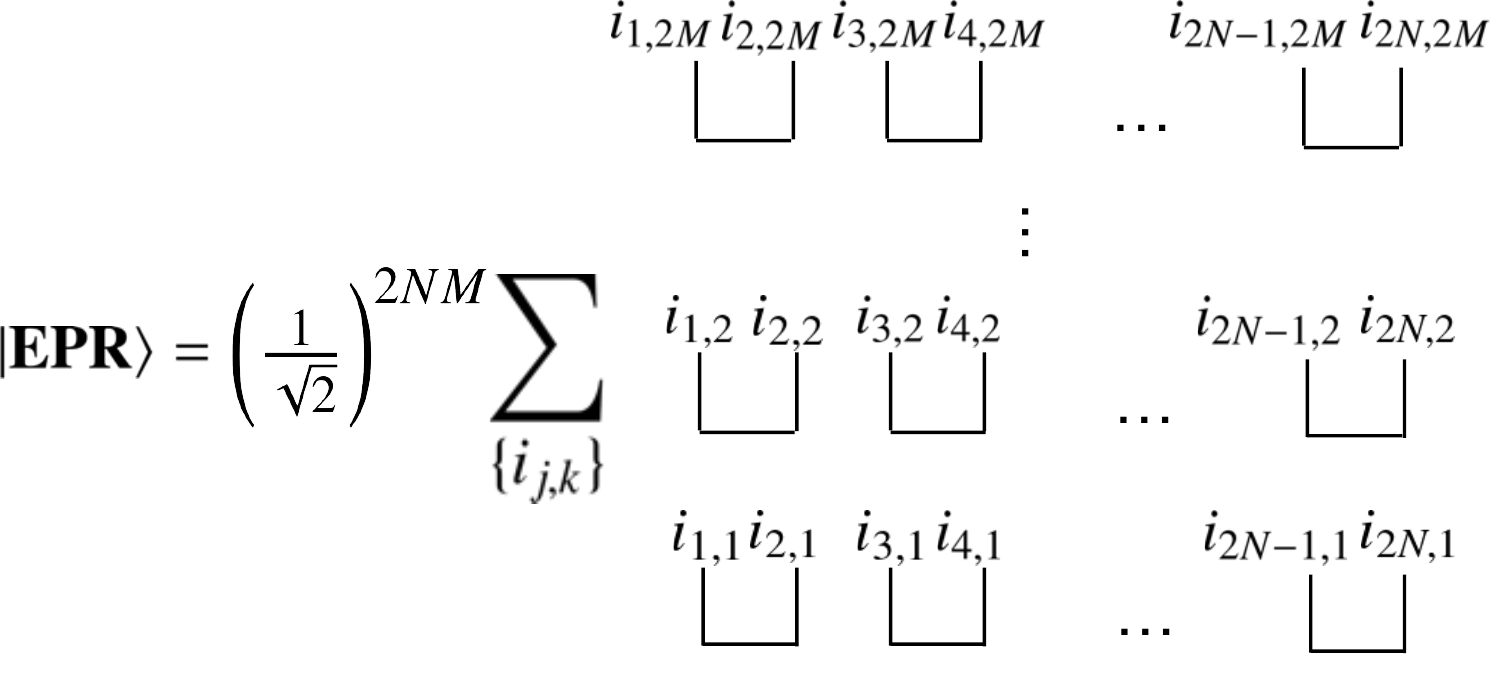}.
\end{equation}

For concreteness,  in this section, we fix $U^{\{2,4\}}(\tau+3)$ and $U^{\{2,4\}}(\tau+1)$ in Eq. (\ref{2D_dual_unitary}) as $U^{(4)}$ and $U^{(2)}$, respectively, but one can discuss the argument in this section similarly in the case of other $U^{\{2,4\}}(\tau+3)$ and $U^{\{2,4\}}(\tau+1)$. 
Besides, as with 1D cases, dynamics $\bm{U_{\rm{2D}}}$ also contain a 2D periodically-driven XXZ  and a 2D self-dual kicked Ising model (The detail is  shown in  Appendix \ref{sec:appC}.).

\subsection{\label{subsec_local} Local expectation values}

We characterize computational complexity of the problem of {calculating} local expectation values for 2D DUQCs.
We consider a local observable in {an} $l \times l$ square region
\begin{equation}
\bm{O}=\prod_{i,j=0}^{l-1} O_{i_0+i,j_0+j},
\end{equation}
where $O_{i_0+i,j_0+j}$ is an observable of qubit $(i_0+i,j_0+j)$ for some integers $i_0$ and $j_0$. Hereafter, for clarity, we assume $l$, $i_0$, and $j_0$ to be odd number, even number, and even number, respectively. 
{Note that the discussion of the following subsections are not limited to this choce, and similar argument can be applied to other choice of parities.}
We denote by $\ket{\mathbf{\Psi}_A(t)}$ a 2D solvable initial state Eq. (\ref{2D_sol_state}) evolved at time $t$, that is, $\ket{\mathbf{\Psi}_A(t)}=\bm{U_{\rm{2D}}}(t)\ket{\mathbf{\Psi}_A}$.
Then, we define the following decision problems as with the 1D case.

\begin{dfn}[local expectation values of 2D DUQCs] \label{2DLE}
Consider a 2D DUQC in time $t$  $\bm{U_{\rm{2D}}}(t)$ with time t,
a 2D solvable $2N \times 2M-$qubit initial state $\ket{\mathbf{\Psi}_A(t)}$ with $M=O$(poly($N$)), 
and a local observable $O$ whose operator norm is 1 with length $l = O(1)$.
$\braket{\mathbf{\Psi}_A(t)|\bm{O}|\mathbf{\Psi}_A(t)}$ is promised to be either $\geq a$ or  $\leq b$, where $a-b \geq \frac{1}{poly(N)}$ 
The problem is to determine whether $\braket{\mathbf{\Psi}_A(t)|\bm{O}|\mathbf{\Psi}_A(t)}$ is $\geq a$ or  $\leq b$.
\end{dfn}

%\begin{problem}
%  \problemtitle{\textsc{local expectation values of 2D DUQCs {at an early time}}}
%  \probleminput{a DUQC $\bm{U_{\rm{2D}}}(t)$ with $t\leq 2N+1-l$, 
%  \newline 2D solvable $2N \times 2M-$qubit initial state $\ket{\mathbf{\Psi}_A(t)}$,
%  \newline local observable $\bm{O}$ whose operator norm is 1
%  \newline with length $l = O(1)$.}
%  \problemquestion{$\braket{\mathbf{\Psi}_A(t)|\bm{O}|\mathbf{\Psi}_A(t)}$ is promised to be either $\geq \frac{2}{3}$ or  %$\leq \frac{1}{3}$. 
%  \newline Determine whether $\braket{\mathbf{\Psi}_A(t)|\bm{O}|\mathbf{\Psi}_A(t)}$ is $\geq \frac{2}{3}$ .}
%\end{problem}

%\begin{problem}
%  \problemtitle{\textsc{local observables of 2D DUQCs in poly time}}
%  \probleminput{a DUQC $\bm{U_{\rm{2D}}}(t)$ with $t=\rm{poly}(N)$, 
%  \newline 2D solvable $2N \times 2M-$qubit initial state $\ket{\braket{\mathbf{\Psi}_A(t)}}$,
%  \newline local observable $\bm{O}$ whose operator norm is 1 
%  \newline with length $l = O(1)$.}
%  \problemquestion{$\braket{\mathbf{\Psi}_A(t)|\bm{O}|\mathbf{\Psi}_A(t)}$ is promised to be either $\geq \frac{2}{3}$ or %$\leq \frac{1}{3}$.  
%  \newline Determine whether  $\braket{\mathbf{\Psi}_A(t)|\bm{O}|\mathbf{\Psi}_A(t)}$ is $\geq \frac{2}{3}$ .}
%\end{problem}

{Similary to the 1D case, we show that  {\bf{Problem} \ref{2DLE}} is in $\sf{P}$ with time $t \leq \lfloor 2(1-\delta)N \rfloor-l$, for $0< \delta <1$, and $\sf{BQP}$-complete with time $t={\rm{poly}}(N)$.}

\subsubsection{{2D DUQC is classically simulatable at an early time}}

\begin{figure*} [tb]
    \centering
    \includegraphics[width=11cm]{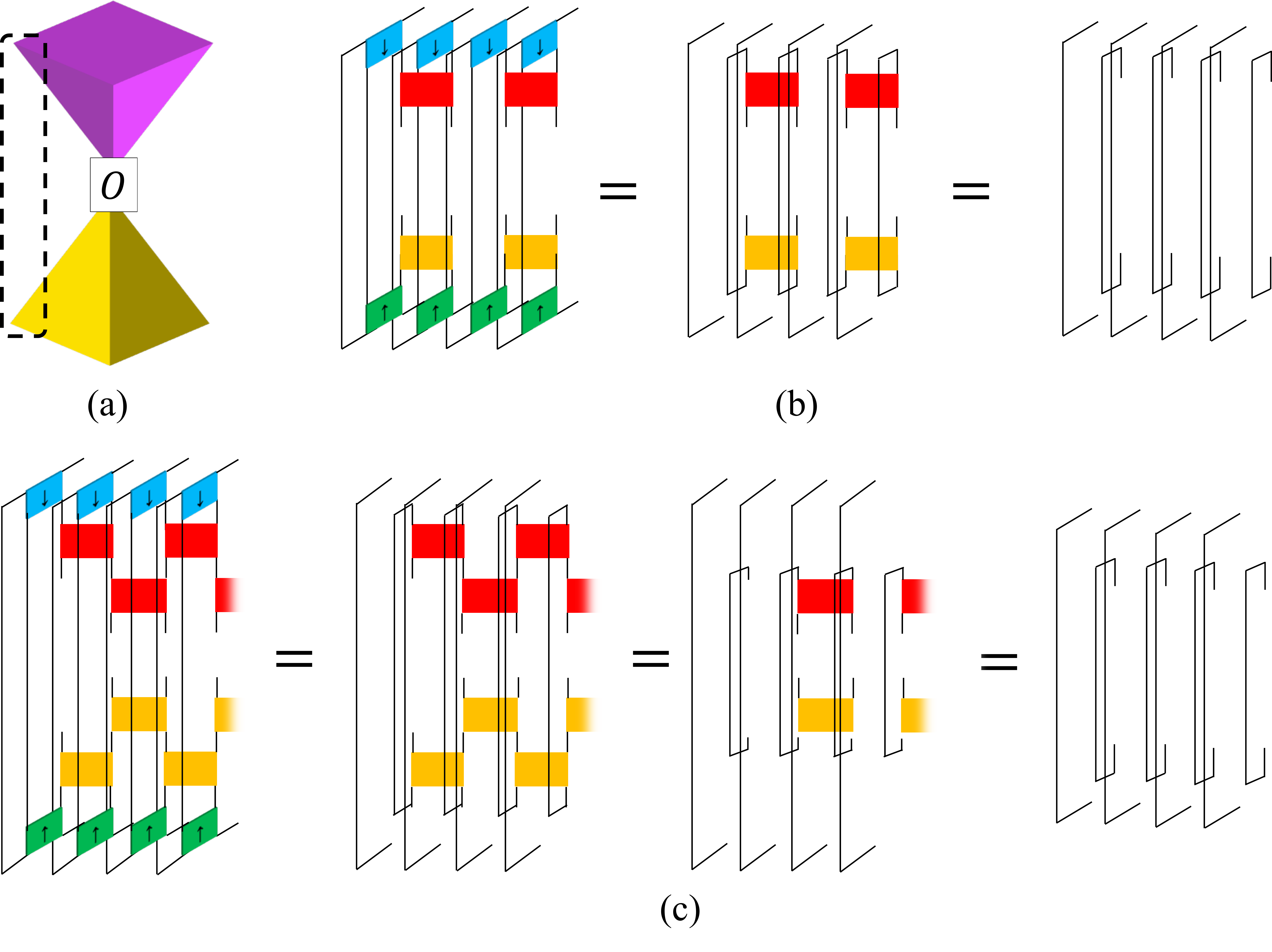}
    \caption{Calculation of a local observable $O$ in a 2D DUQC . After removing unitary gates outside of the causal-cone, remaining unitary gates can be represented as (a). Contractions of dual-unitary gates and unitary gates in the dotted line of (a) are represented in (b). Here, a unitary gate painted in red is a harmitian conjugate of one painted in orange. Contractions of dual-unitary gates and unitary gates, in the case that $U^{(2)}$ of (b) are replaced by $U^{(4)}U^{(2)}$, are represented in (c).}
    \label{fig:2dobs}
\end{figure*}

The way to calculate local expectation values is similar to the 1D case except for contractions of unitary gates at even time. 
The procedure is to contract {dual-unitary gates} at odd time by using Eq. (\ref{ucont}) and unitary gates at even time by using Eq. (\ref{ducont}). As a result, if $t \leq \lfloor 2(1-\delta)N \rfloor-l$ and $t \geq l+1$, we obtain 
\begin{equation}
    \braket{{\mathbf{\Psi}}_{t}|\bm{O}|{\mathbf{\Psi}}_{t}}=\frac{1}{2^{l^2}}{\rm{Tr}}(O) + O \left( M \cdot  | \lambda _1|^{\lfloor \delta N \rfloor} \right).
\end{equation}
Origins of the conditions $t \geq l$ are the same as those of 
1D cases.
We explain the procedure graphically in the case that an initial state is rows of EPR pairs (Fig. \ref{fig:2dobs}).

Firstly, we remove unitary gates out-side of the causal-cone.
Then, {the} local expectation value can be represented as Fig. \ref{fig:2dobs} (a).
A tensor network in dotted line of Fig. \ref{fig:2dobs} (a) is depicted in further detail in Fig. \ref{fig:2dobs} (b).
In the first {equality} in Fig. \ref{fig:2dobs} (b), we use the definition of dual-unitary gates Eq. (\ref{ducont}), and remove them.
After that, unitary gates, which are contracted with removed dual-unitary gates (unitary gates painted in red and orange in Fig. \ref{fig:2dobs}), can be removed by the definition of unitary gates  Eq. (\ref{ucont}).
This leads to the second equation in (b).
By repeating this procedure, one obtain that local expectation values are ${\rm{Tr}}(O)$.
Moreover, local expectation values {at time less than $l+1$} are classically simulatable because only constant number of unitary gates are involved in calculation of local expectation values. 
One can straightforwardly generalize the above results to general 2D solvable initial states {similar to} the 1D case and Appendix \ref{sec:appB}. 
Altogether, we obtain {\bf{Problem} \ref{2DLE}} with $t \leq \lfloor 2(1-\delta)N \rfloor-l$ is in $\sf{P}$. 
{Note that this is true even if $U^{(2)}$ and $U^{(4)}$ are replaced by $U^{\{2,4\}}$. For example, in the case of $U^{\{2,4\}}=U^{(4)}U^{(2)}$ , dual-unitary gates and unitary gates are contracted and removed as Fig. \ref{fig:2dobs} (c).}

{We note that classical simulatability of 2D DUQC are also interesting as a solvable model of quantum dynamics because analytical research on quantum dynamics in 2D systems is much more difficult than 1D cases. 
In fact, all local expectation values of 2D DUQCs become Tr($O$) in the thermodynamic limit, which means that a local density matrix of a 2D solvable initial state evolved by 2D DUQCs is identical to a thermal equilibrium state at infinite temperature.
Therefore, thermalization of solvable initial states can be shown analytically in the thermodynamic limit.
Understanding conditions when thermalization happens is one of the most important problems in non-equilibrium physics \cite{RevModPhys.83.863, eisert2015quantum, d2016quantum, mori2018thermalization}, and it means that 2D DUQCs are rare toy models of 2D non-equilibrium quantum physics.}

\subsubsection{{2D DUQC is universal in late time}}

Based on the fact that {\bf{Problem} \ref{1DLE}} with time $t={\rm{poly}}(N)$ is $\sf{BQP}$-complete , $\sf{BQP}$-completeness of {\bf{Problem} \ref{2DLE}} with time $t={\rm{poly}}(N)$ becomes trivial by noticing that $\bm{U_{\rm{2D}}}(t)$ acting on any row of $2N$-qubit can be regarded as 1D DUQCs when both $U^{(2)}$ and $U^{(4)}$ are identity operators. 

%As one would expect, if time t is sufficiently large polynomial in $N$, 2D DUQCs can perform universal quantum computation. This follows from the fact that 1D DUQCs can perform universal quantum computation. Therefore, in worst case, these dynamics are unlikely to be simulated by a classical computer, as with  1D cases.

\subsection{Sampling problem}
We now move on to discuss the sampling complexity of 2D DUQCs. 
We show that a constant depth 2D DUQC can generate a square lattice cluster state.

We obtain a square lattice cluster state $\ket{\rm{CS}}$ by four depth 2D DUQCs as follows:
\begin{align}
& \ket{{\mathbf{\Psi}}_{1}}=\prod_{i=1}^{2N}\prod_{j=1}^{N}({\rm{SWAP}}\cdot {\rm{CZ}} \cdot  H \otimes I)_{(i,2j),(i,2j+1)}\ket{\mathbf{EPR}},\\
& \ket{{\mathbf{\Psi}}_{2}}=\prod_{i=1}^{N}\prod_{j=1}^{2N}{\rm{CZ}}_{(2i-1,j),(2i,j)}\ket{{\mathbf{\Psi}}_{1}},\\
& \ket{{\mathbf{\Psi}}_{3}}=\prod_{i=1}^{2N}\prod_{j=1}^{N}{\rm{SWAP}}_{(i,2j-1),(i,2j)}\ket{{\mathbf{\Psi}}_{2}},\\
& \ket{\rm{CS}}=\prod_{i=1}^{N}\prod_{j=1}^{2N}{\rm{CZ}}_{(2i,j),(2i+1,j)}\ket{{\mathbf{\Psi}}_{3}}.
\end{align}
Therefore, sampling the output of constant depth 2D DUQCs is unlikely to be classically simulatable.

\subsection{Correlation functions} \label{subsec_Correlation functions}

Let us discuss classical simulatability of two-point correlation functions. For simplicity, we assume that the initial state is $\ket{\mathbf{EPR}}$, but one can easily generalize the following argument to general solvable initial states. 
In such a case, we expect two-point correlation functions to be anisotropic because solvable initial states are anisotropic.
We consider two {types of}  correlation functions, one of which is classically simulatable, and {the other} does not seem to be classically simulatable.

\begin{figure*} [tb]
    \centering
    \includegraphics[width=9cm]{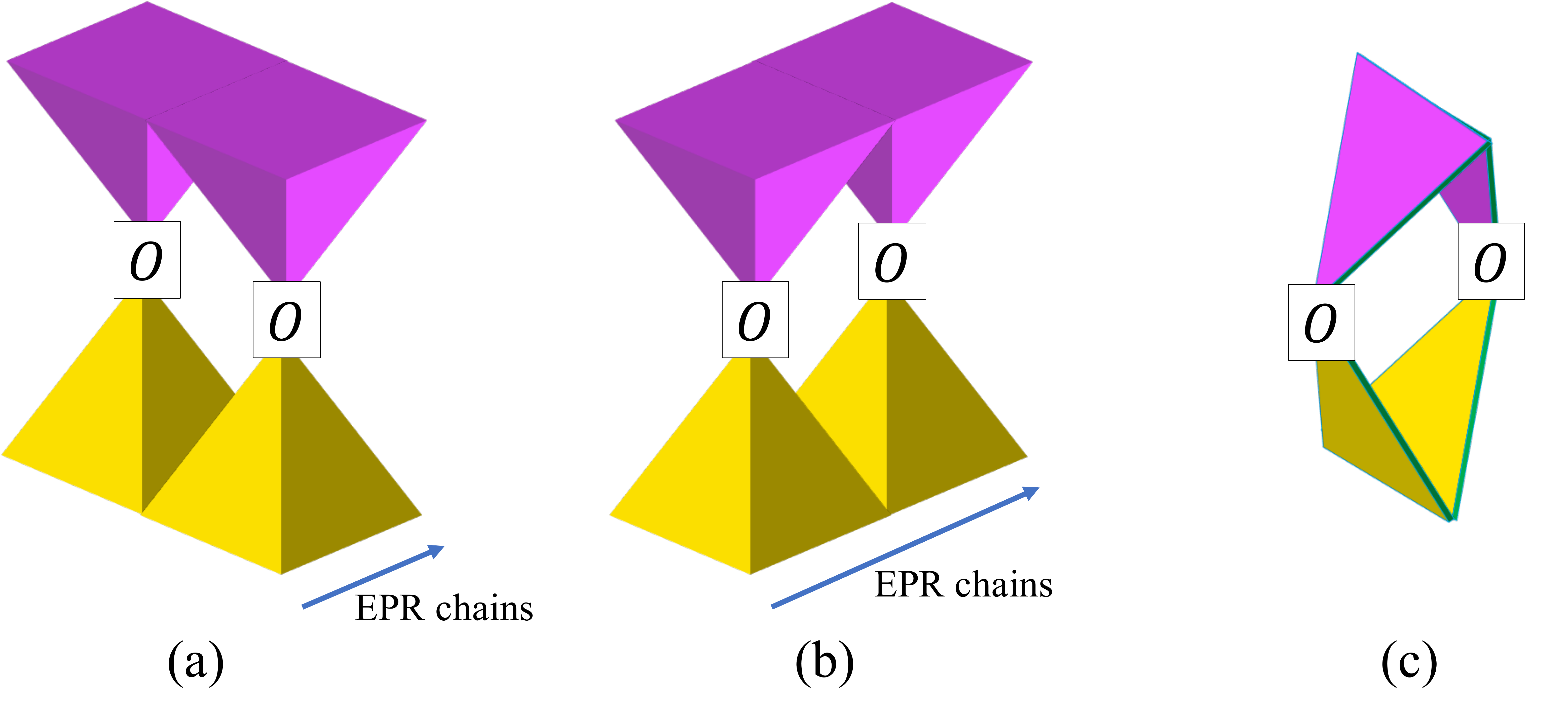}
    \caption{Calculation of correlation functions. 
    After removing unitary gates outside of the causal-cone, remaining unitary gates of $C_1(r,t)$ and $C_2(r,t)$ can be represented as (a) and (b), respectively, in the case that $C_2(r,t)$ can take a nonzero value.
    EPR chains in (a) and (b) are aligned along arrows in (a) and (b), respectively.
    After removing dual-unitary at odd time by using Eq. (\ref{ucont}) and unitary gates at even time by using Eq. (\ref{ducont}), remaining unitary gates of $C_2(r,t)$ can be represented as (c).}
    \label{fig:correlation}
\end{figure*}

First, we consider {the} following correlation function, which is classically simulatable:
\begin{align}
C_1(r,t)=\braket{{\mathbf{EPR}_t}|O_{i,j}O_{i+r,j}|{{\mathbf{EPR}_t}}}-{\rm{Tr}}(O_{i,j}){\rm{Tr}}(O_{i+r,j}), \label{cf_1}
\end{align}
where $\ket{\mathbf{EPR}_t}$ is $\ket{\mathbf{EPR}}$ evolved at time $t$, and $O_{i,j}$ is {an} observable of qubit ($i,j$). 
When $2N \geq 2t-r$ holds, $C_1(r,t)$, which can be graphically represented as Fig. \ref{fig:correlation} (a), can be straightforwardly calculated as with calculation of expectation values shown in Fig. \ref{fig:2dobs}{. As a result,} we have
    $C_1(r,t)=0.$
{In other words, qubit $(i,j)$ and qubit $(i+r,j)$ cannot be correlated for an arbitrary 2D DUQC in the time region.}
Second, we consider {the} following correlation function, which is {expected to be classically intractable}:
\begin{align}
C_2(r,t)=\braket{{\mathbf{EPR}}_t|O_{i,j}O_{i,j+r}|{{\mathbf{EPR}_t}}}-{\rm{Tr}}(O_{i,j}){\rm{Tr}}(O_{i,j+r}) \label{cf_2}.
\end{align}
We also assume that $2N \geq 2t-r$ holds. 
{$C_2(r,t)$ can take a nonzero value only in the case of $r=t+1$ and odd $j$, $r=t$, or $r=t-1$ and even $j$.
In those cases, we conjecture that $C_2(r,t)$ is unlikely to be classically simulatable because of the following reason.}
$C_2(r,t)$ can be graphically represented as Fig. \ref{fig:correlation} (b).
 After {contracting} unitary and dual-unitary gates of Fig. \ref{fig:correlation} (b) by using Eq. (\ref{ucont}) and Eq. (\ref{ducont}), the remaining gates of it can be represented as Fig. \ref{fig:correlation} (c).
This is similar to correlation functions of 1D DUQCs (see Fig. 3 of Ref. \cite{PhysRevB.101.094304}), but important difference is that uncontracted gates form a 2D tensor-network in 2D DUQCs. 
Because of this, calculating correlation functions in 2D DUQCs seems to be hard for a classical computer. 
Therefore, classical simulatability of correlation functions seems to depend on {the relative position of two local observables.}
It is still an open problem whether or not calculating $C_2(t,t)$ with the condition $2N \geq 2t-r$ is $\sf{BQP}$-hard.

\subsection{General lattice pattern}

So far, we only consider dynamics in a 2D square lattice. 
It is natural to consider a generalization to other lattices, such as a honeycomb lattice. 
We note that unitaries at even-time of dynamics defined in Sec. \ref{subsec_def} can be chosen as identity gates. 

\begin{figure*} [tb]
    \centering
    \includegraphics[width=15cm]{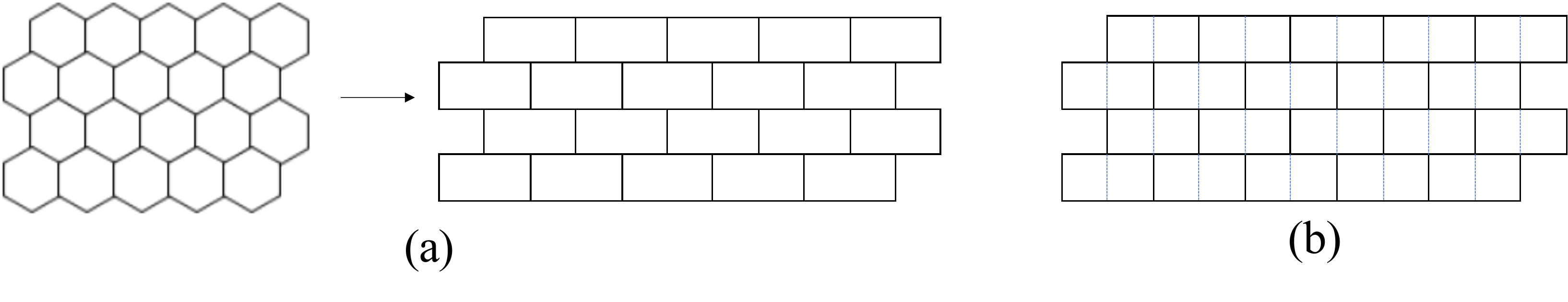}
    \caption{Equality of a honeycomb lattice and a square lattice up to longitudinal edges. As illustrated by (a), a honeycomb lattice can be deformed into a lattice which is made by rectangles. This deformed lattice is equal to a square lattice up to longitudinal edges, as illustrated by (b).}
    \label{fig:honeycomb}
\end{figure*}

If a unitary gate $U_{(j,k),(j,k+1)}$ is an identity gate at all time, the edge between qubit $(j,k)$ and qubit $(j,k+1)$ {can be effectively removed}. 
So, if we construct lattices by eliminating edges in $k$-direction, 2D DUQCs with such a lattice can be treated as with ones with a 2D square lattice. 
We name such lattices solvable lattices. 
Such lattices include, for example, a honeycomb lattice. This is illustrated by Fig. \ref{fig:honeycomb}. 
One can construct an arbitrary solvable lattice in the same way.
{Therefore, local expectation values and correlation functions $C_1$ of 2D DUQCs with solvable lattices  are classically simulatable at an early time as same as those with square lattices.}

\section{\label{sec:Conclusion} Conclusion and discussion}
We have investigated computational complexity of the problems calculating physical properties of 1D and 2D  DUQCs.
First, we have shown {that} the complexity of calculating local expectation values of dual-unitary quantum circuits highly depends on their circuit depth.
Second, we have shown that classical simulation of sampling from 1D and 2D  DUQCs after linear and constant depth, respectively, is hard.

The first result is in contrast to conventional classically simulatable  quantum circuits with fixed gate sets such as Clifford circuits and matchgates, whose {classical simulatability does not change depending on the circuit depth}.
The dual-unitary quantum computational model is the first example of the model, where quantum computational power makes a transition between $O$($N$) time and poly($N$) time.
It is reminiscent of dynamical phase transitions of computational complexity which have been recently studied in other contexts \cite{deshpande2018dynamical, maskara2020complexity, napp2020efficient}.
It would be interesting to investigate whether or not local expectation values of DUQCs {at an early time slightly later} than that considered in this paper are also classically simulatable.
{Another future direction would be to study the computational power of DUQCs in linear depth with non-solvable initial states. In this context, it is well known that Clifford circuits and matchgates circuits with  certain initial states, so-called magic states, have potential to outperform those without magic states \cite{bravyi2005universal, hebenstreit2019all}. Thus, it is natural to expect that there exist magic states enhancing the computational power of DUQCs at an early time, and we leave it for future work. }

{In the second result, we have considered classical sampling from a DUQC with a multiplicative error. Here, we note that  classical simulation of sampling with an additive error from certain quantum computing models, such as linear optical circuits \cite{10.1145/1993636.1993682}, IQP circuits \cite{bremner2016average}, and the DQC1 model \cite{morimae2017hardness}, is known to be hard under plausible assumptions. 
With this in mind, because of the computational universality of DUQCs, sampling with an additive error from DUQCs which simulates IQP circuits is also intractable for classical computers. 
Another sampling problem, which attracts much attention from both theorists and experimentalists, is random circuit sampling \cite{bouland2019complexity, arute2019quantum}.
It would be interesting to investigate whether or not classical sampling with an additive error from a random DUQC, where every gate is chosen randomly from the set of dual-unitary gates, is hard.}

{Moreover, we remark that our argument on the computational universality and sampling complexity of one- and two-dimensional DUQCs can be straightforwardly extended to the case of open boundary conditions (OBCs). However, under the condition, classical simulation of DUQCs at an early time becomes harder since, in general, unitary gates at the boundary of causal-cone cannot be cancelled. We discuss this point in some more detail in Appendix \ref{sec:appD}, and we leave a full characterization of their computational power for future work.
}

Besides, because analytic research on quantum dynamics in 2D systems is much more difficult than 1D cases, generalization of DUQC to two spatial dimensions are also interesting as a solvable model of quantum dynamics. 
It would be important to generalize DUQCs to higher than two spatial dimensions {and construct more general solvable initial states, for example, using higher-dimensional tensor-network states such as projected entangled pair states (PEPSs) \cite{RevModPhys.93.045003} }.
We expect that a high-dimensional DUQC will deepen our understanding of a non-equilibrium phenomenon in a high-dimensional quantum system, such as a 2D self-dual kicked Ising model.

{
\section{\label{sec:acknowledgement} ACKNOWLEDGMENTS}
This work is supported by
MEXT Quantum Leap Flagship Program (MEXT QLEAP) Grant Number JPMXS0118067394 and JPMXS0120319794. 
KM is supported by JST PRESTO Grant No. JPMJPR2019 and JSPS KAKENHI Grant No. 20K22330. KF is supported by JSPS KAKENHI Grant No. 16H02211, JST ERATO JPMJER1601, and JST CREST JPMJCR1673. 
}

\appendix
\section{A calculation procedure for local expectation values of a chain of EPR pairs in 1D DUQCs \label{sec:appA}} 

In this appendix, we show that local expectation values for $2N \geq l+2(t-1)$:
\begin{equation} \label{Evalue1:app}
\includegraphics[clip,width=7cm]{Evalue1n.pdf}
\end{equation}
is equal to $\frac{1}{2^l}{\rm{Tr}}(O)$. 
The procedure is similar to that in Refs. \cite{gopalakrishnan2019unitary, PhysRevLett.123.210601, PhysRevB.101.094304}. 

To begin with, Eq. (\ref{Evalue1:app}) is equal to
\begin{equation} \label{Evalue1_22:app}
\includegraphics[clip,width=5cm]{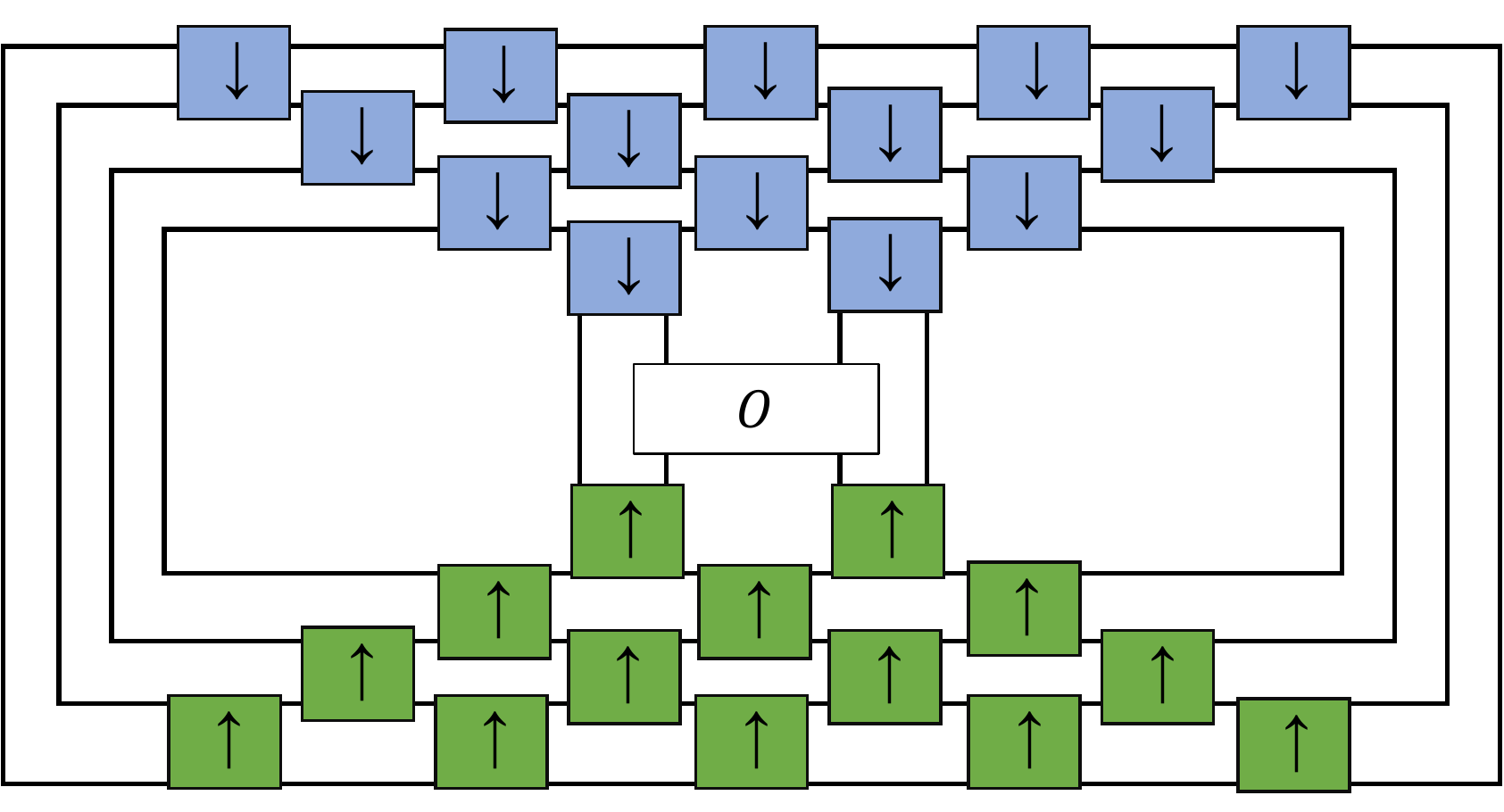}.
\end{equation}
By adapting Eq. (\ref{ducont}) to the leftmost and rightmost unitaries, Eq. (\ref{Evalue1_22:app}) becomes 
\begin{equation} \label{Evalue1_3:app}
\includegraphics[clip,width=4.5cm]{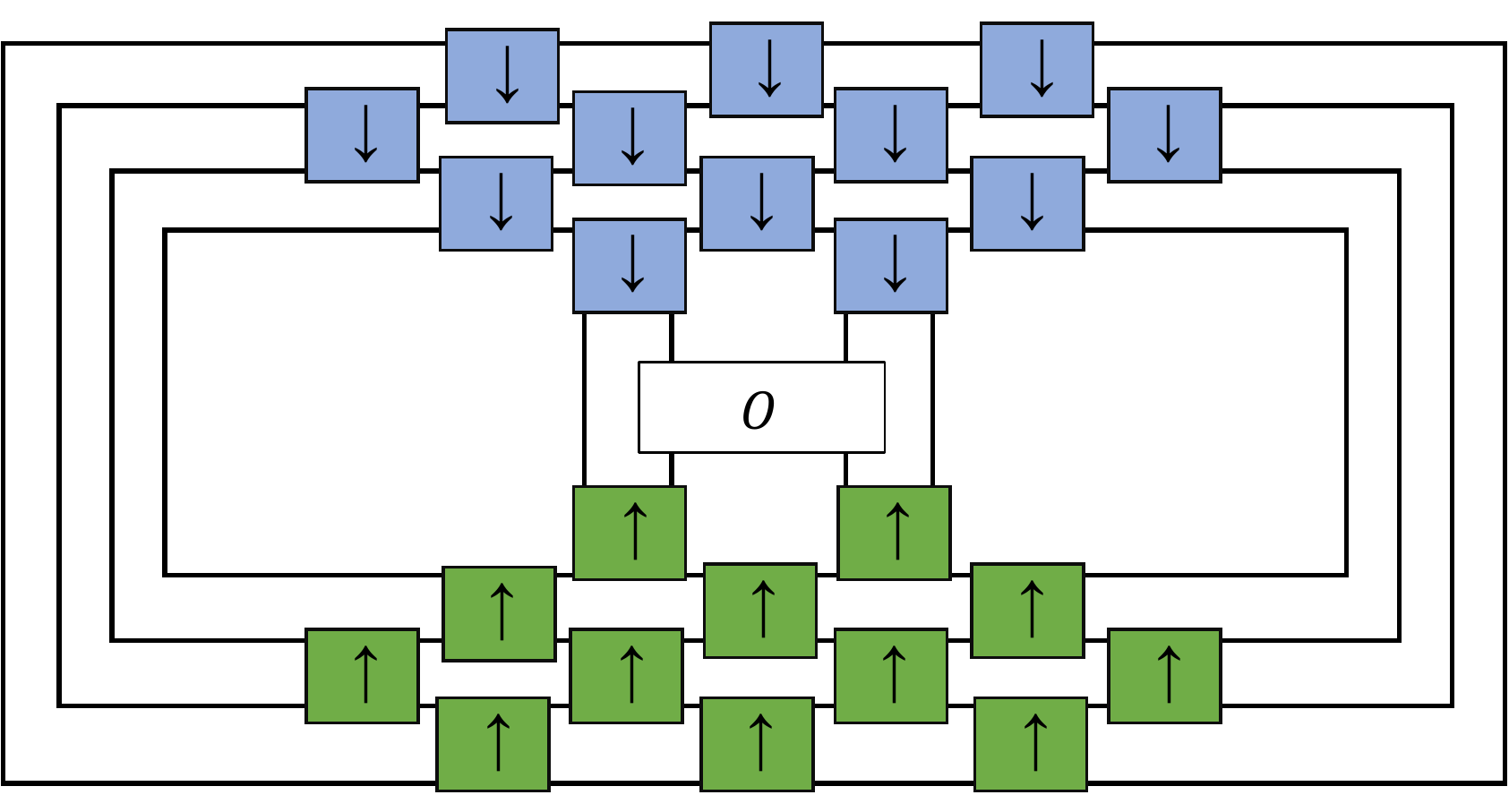}.
\end{equation} 
By repeating the process, Eq. (\ref{Evalue1_3:app}) becomes 
\begin{equation} \label{Evalue1_4:app}
\includegraphics[clip,width=5cm]{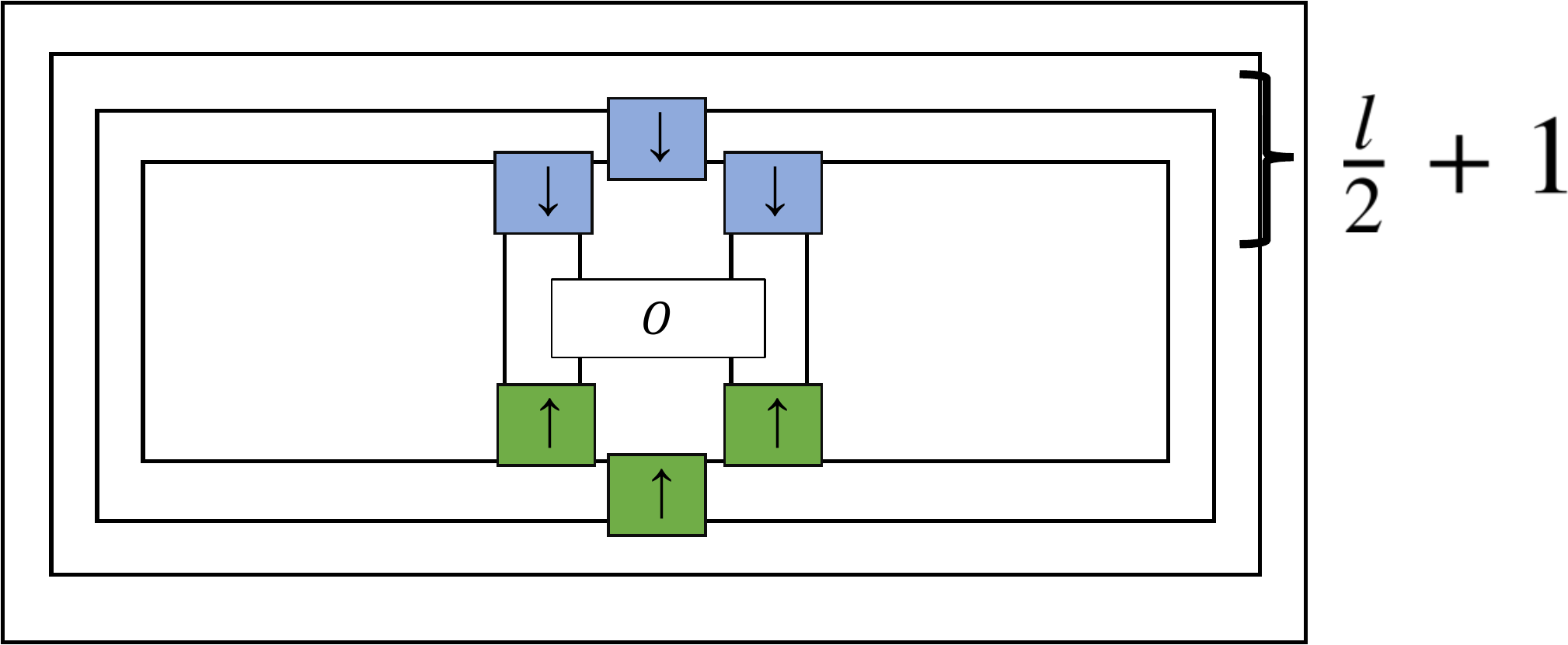}.
\end{equation} 
Finally, by using Eq. (\ref{ucont}) repeatedly, Eq. (\ref{Evalue1_4:app}) becomes 
\begin{equation} \label{Evalue1_5:app}
\includegraphics[clip,width=1.8cm]{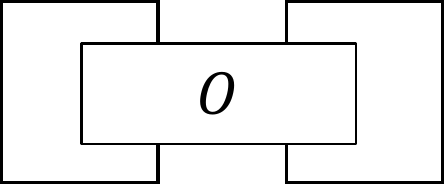}.
\end{equation} 
{Taking} the renormalization coefficient into account, one obtains $\braket{\Psi^N_t|O|\Psi^N_t}=\frac{1}{2^l}{\rm{Tr}}(O)$.

\section{Local expectation values of general solvable initial states.  \label{sec:appB}} 
In this appendix, we show that local expectation values at time ${t}$ with general solvable states are approximated by $\frac{1}{2^l}$Tr($O$). 
Local expectation values without normalization can be written as
\begin{equation} \label{error1.0}
\includegraphics[clip,width=6.5cm]{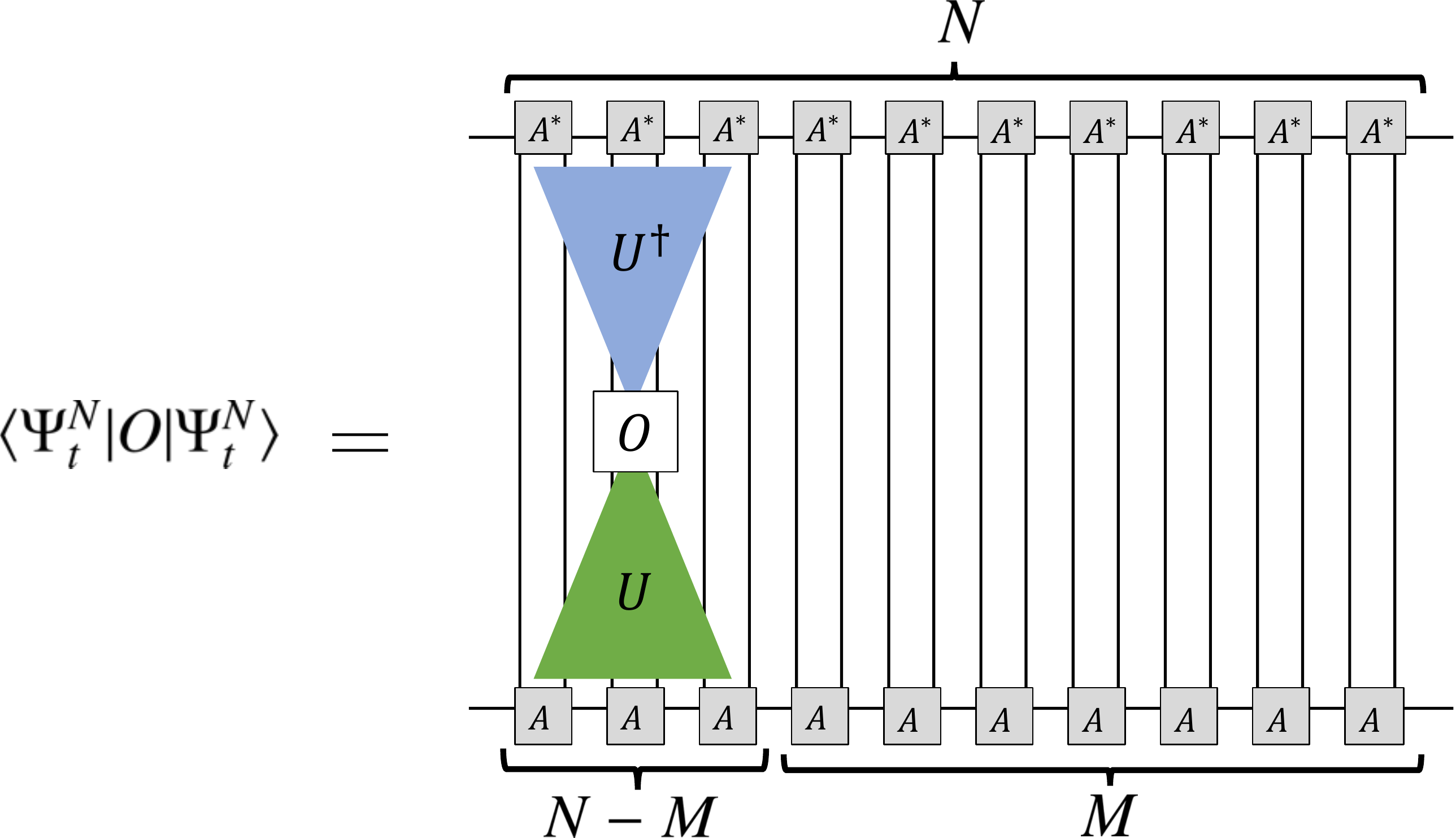},
\end{equation}
where $N$ and $M$ are {the total number of tensors $A$ and ones} which are outside of the causal-cone, respectively.
By inserting identity operators $I=\sum_{\alpha=1}^{\chi}\ket{\alpha}\bra{\alpha}$, one obtains
\begin{equation} \label{error1.1}
\includegraphics[clip,width=8cm]{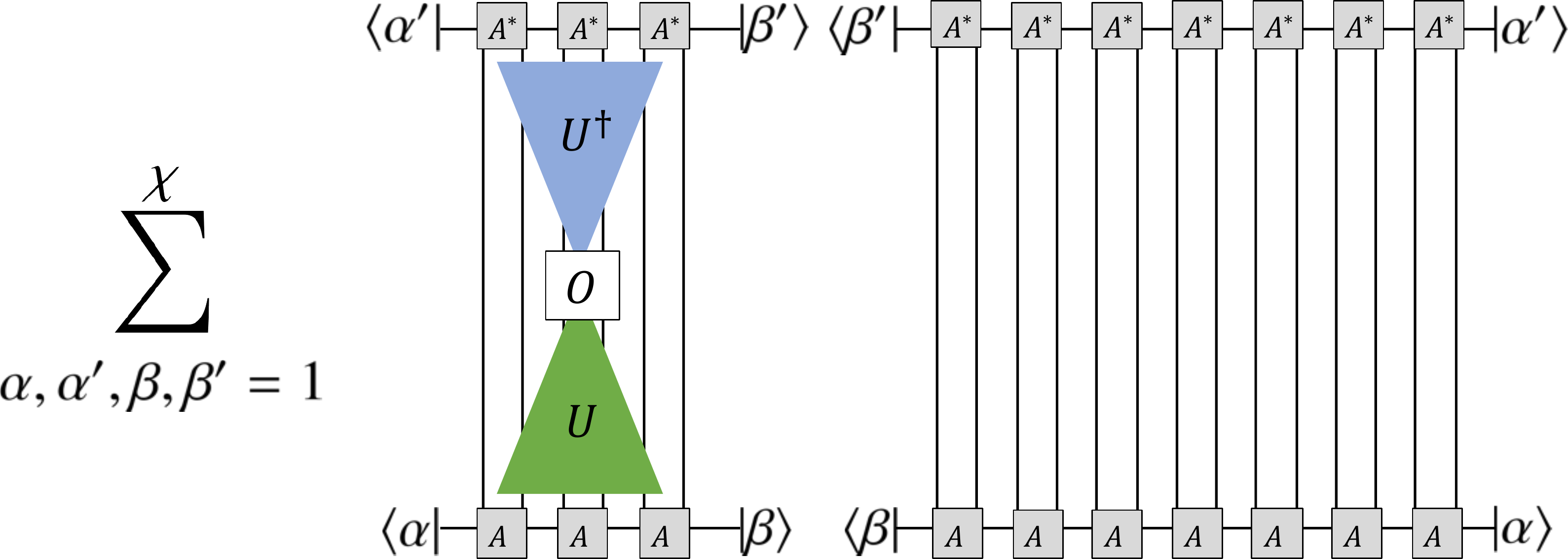}.
\end{equation}
{The second factor of Eq. (\ref{error1.1}) is equal to a matrix element of $E^M$, where $E$ is the transfer matrix defined in Eq. (\ref{transfer})}.
{By the assumption that $\ket{I}$ is the unique eigenvector of the matrix $E$ with the latgest eigenvalue $1$, the Jordan canonical form of $E$ can be written as}
\begin{align}
    E= \ket{I}\bra{I}+ S \left( \sum_{i=1}^{{D}}(\lambda_i P_i + N_i) \right) S^{-1},
\end{align}
where {$D$ is the number of Jordan blocks}, $P_i$ is the diagonal part, $N_i$ is the nilpotent part, and $\lambda_i$ is less than $1$ and ordered in descending order.
Let $\varepsilon$ be $\sum_i(\lambda_i P_i + N_i)$.
Then, the n-th power of a transfer matrix $E^M$ can be written as $E^M= \ket{I}\bra{I}+ S \varepsilon^M  S^{-1}$.

Now we evaluate local expectation values. 
First, in the same way as the calculation in Appendix \ref{sec:appA}, we obtain that 
\begin{equation} \label{error1.2}
\includegraphics[clip,width=1.8cm]{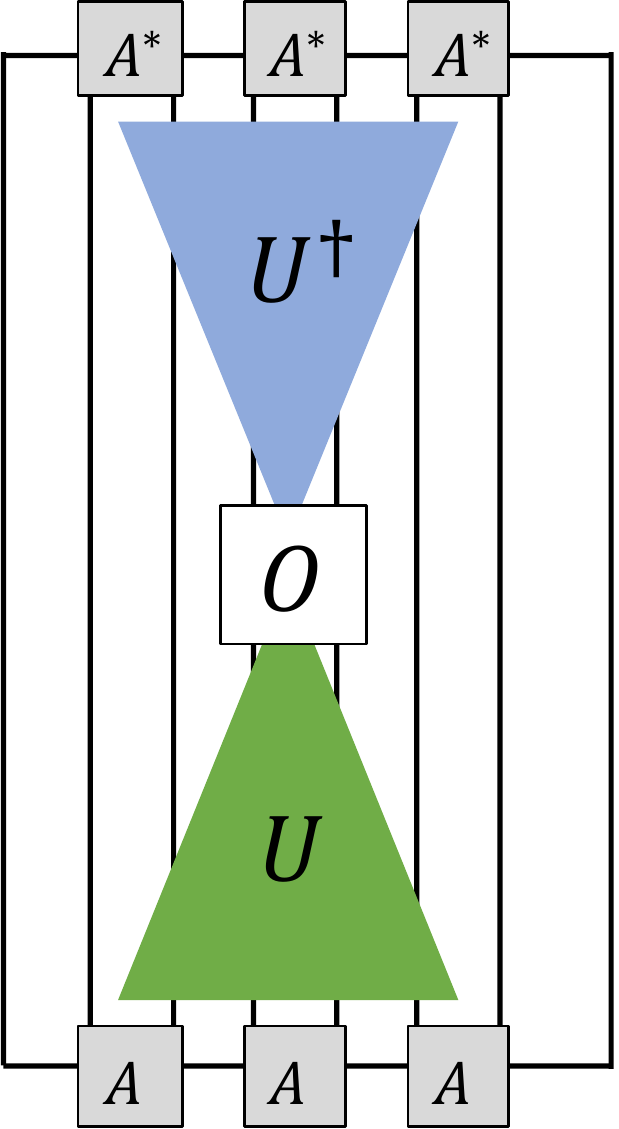}
\end{equation}
is equal to $\frac{1}{2^l}$Tr($O$), which arises from $\ket{I}\bra{I}$ and is a dominant term of Eq. (\ref{error1.1}).

Next, we calculate the error term $\epsilon = \braket{\Psi^N_t|O|\Psi^N_t}-\frac{1}{2^l}{\rm{Tr}}(O)$, which arises from $S \varepsilon^M S^{-1}$.
First, we evaluate the second factor of Eq. (\ref{error1.1}), namely, $\braket{\beta' \beta|S \varepsilon^M  S^{-1}| \alpha' \alpha}$. 
{We denote the $L_2$-norm of the vector $\ket{a}$ by $\|\ket{a}\|$ and the $L_2$ operator norm of a matrix $B$ by $\|B\|$.
Because of
\begin{equation}
\max_{\|\ket{x}\|=\|\ket{y}\|=1}{|\braket{y|B|x}|}=\|B\|, \label{eq:max}
\end{equation}}
where $B$ is a matrix, the following holds:
\begin{equation} \label{eq:error1}
\braket{\beta' \beta|S \varepsilon^M S^{-1}| \alpha' \alpha} \leq \|S\| \cdot \|S^{-1}\| \cdot \|\varepsilon^M\|.
\end{equation}
By the binomial theorem, $\varepsilon^M$ is expanded as 
\begin{align}\label{epsilon}
\varepsilon^M=\sum_{i=1}^{{D}} \left( \sum_{j=0}^{d_i-1} {}_M C _j \lambda_i^{M-j} P_iN_i^j \right),
\end{align}
where $d_i$ is the dimension of the $i$-th Jordan block, that is, the dimension of the space on which  $P_i$ acts nontrivially, and we used the fact that $N_i^{d_i}$ is the zero matrix.
Then, $\| \varepsilon^M\|$ is upper bounded as follows:
\begin{align}
\| \varepsilon^M\| &\leq \sqrt{\sum_{\alpha,  \beta=1}^{\chi^2}\left| \left(\varepsilon^M \right) _{\alpha \beta} \right| ^2} \\
&\leq \sqrt{\chi^4 \max_{i,j} \left( \left( {}_M C_{j} \right)^2 \lambda_i^{2(M-j)}\right) }\\
&= \chi^2 \max_{i,j} \left( {}_M C_{j} \lambda_i^{M-j}\right), \label{eq:error2}
\end{align}
for $j=0, 1, \cdots, d_i$,
where {the first {inequality} follows from the fact that for a matrix $B$, $\|B\|$ is less than its Frobenius norm $\|B\|_F=\sqrt{\sum_{i,j=1}B_{ij}^2}$ \ \cite{golub2013matrix},
and the second inequality follows from the fact that the maximum value of matrix elements of $\varepsilon^M$ is $\max_{i,j} \left( {}_M C_{j} \lambda_i^{M-j}\right)$ due to Eq. (\ref{epsilon}) }.
From Eqs. (\ref{eq:error1}) and (\ref{eq:error2}), the following holds:
\begin{align}
\braket{\beta' \beta|S \varepsilon^M S^{-1}| \alpha' \alpha} = O(\lambda_1^{M}), \label{eq:order1}
\end{align}
where we used the inequality $d_i < \chi^2$ and the assumption $\chi=O(1)$.

Second, we evaluate the first factor of Eq. (\ref{error1.1}), namely,
\begin{align} \label{eq:GeneObs}
\includegraphics[clip,width=2.5cm]{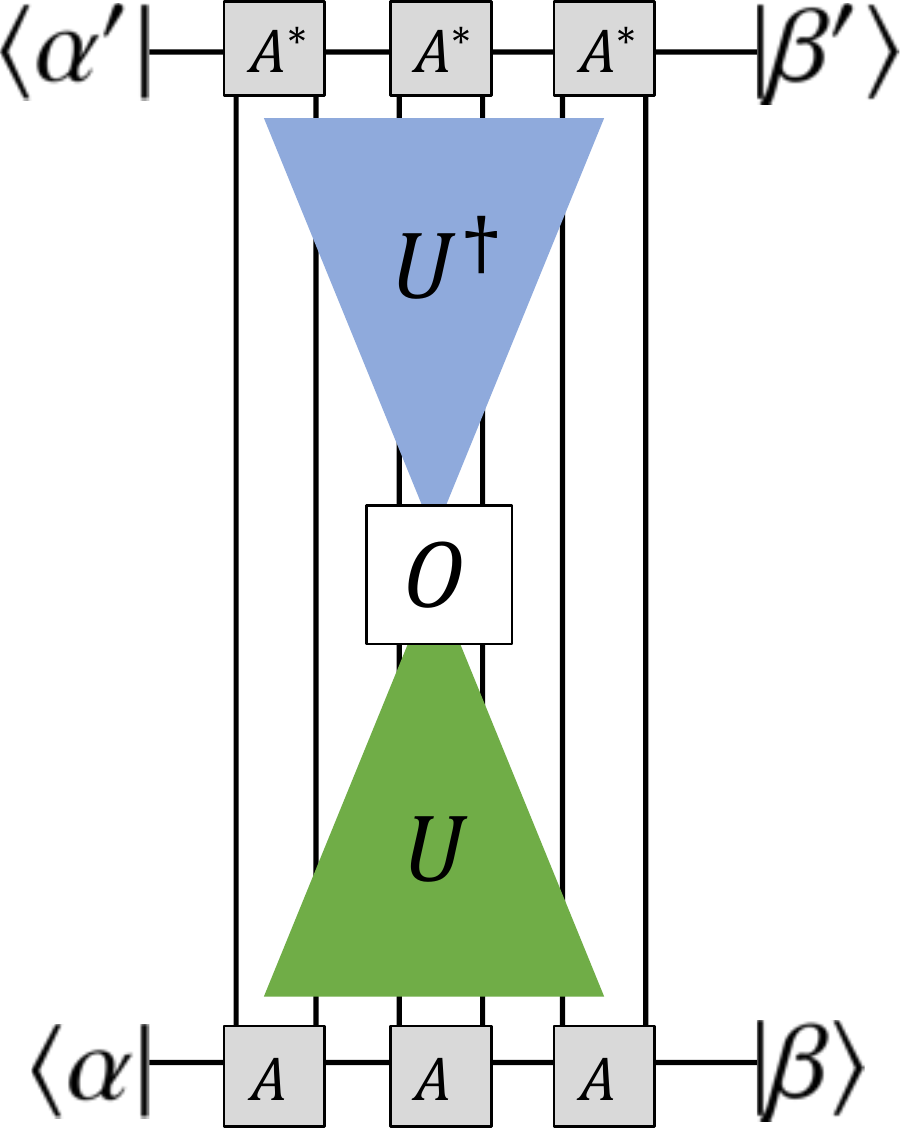}.
\end{align}
Let $\ket{\Psi_{(\alpha, \beta)}}$ be, 
\begin{align} \label{eq:alphabeta}
\includegraphics[clip,width=2.5cm]{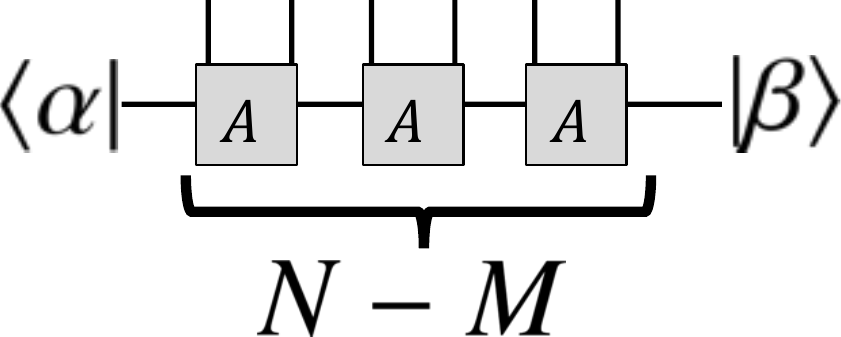}.
\end{align}
Then, Eq. (\ref{eq:GeneObs}) can be written as
\begin{align} \label{eq:GeneObs1}
\braket{\Psi_{(\alpha', \beta')}|U^{\dagger}OU|\Psi_{(\alpha, \beta)}},
\end{align}
and this is upper bounded as follows:
\begin{align} \label{eq:GeneObs2}
&\braket{\Psi_{(\alpha', \beta')}|U^{\dagger}OU|\Psi_{(\alpha, \beta)}} \\
&= \|\Psi_{(\alpha', \beta')}\| \cdot \|\Psi_{(\alpha, \beta)}\| \cdot 
\frac{\bra{\Psi_{(\alpha', \beta')}}}{\|\Psi_{(\alpha', \beta')}\|} U^{\dagger}OU \frac{\ket{\Psi_{(\alpha, \beta)}}}{\|\Psi_{(\alpha, \beta)}\|}\\
& \leq \|\Psi_{(\alpha', \beta')}\| \cdot \|\Psi_{(\alpha', \beta')}\| \\
&= \left( \braket{\alpha' \alpha'|E^{N-M}| \beta' \beta'} \right)^{1/2} \left( \braket{\alpha \alpha|E^{N-M}| \beta \beta}\right)^{1/2} \\
&= \left( \braket{\alpha' \alpha'|(\ket{I}\bra{I}+ S \varepsilon^M  S^{-1})| \beta' \beta'} \right)^{1/2} \left( \braket{\alpha \alpha|( \ket{I}\bra{I}+ S \varepsilon^M  S^{-1})| \beta \beta}\right)^{1/2} \\
& \leq 1+ \chi^2 \max_{i,j} \left( {}_M C_{j} \lambda_i^{M-j}\right)\\
&= O(1) \label{eq:order2}
\end{align}
where in the first inequality, we used Eq. (\ref{eq:max}) and $\| O \|=1$, 
and in the second inequality, we used Eqs. (\ref{eq:error1}), (\ref{eq:error2}), and $\braket{kk|I}=\frac{1}{\chi}$, {for any $k \in \{\alpha, \alpha', \beta, \beta'\}$. }

From Eqs. (\ref{eq:order1}) and (\ref{eq:order2}), the following holds:
\begin{align} \label{eq:order3}
\varepsilon=O(\lambda_1^{M}).
\end{align}
Because of {the condition that} $M=2N-l-2t$ in Eq. (\ref{Evalue3}), 
$\varepsilon$ becomes $O(\lambda_1^{2N-l-2t})$.
In addition, because $\braket{\Psi^N_t|\Psi^N_t}$ is $1+O(\lambda_1^N)$, normalized local expectation values $\braket{O(t)}$ is
\begin{align}
\braket{O(t)}={\rm{Tr}}(O)+O(\lambda_1^{2N-l-2t})+{O(\lambda_1^N)}.
\end{align}
Then, if $t$ satisfies $t \leq \lfloor (1-\delta)N \rfloor - {l/2}$ for some $0< \delta <1$, the error term $O(\lambda_1^{2N-l-2t})$ become $O(\lambda_1^{\lfloor \delta N \rfloor})$. Therefore, local expectation values with $t \leq \lfloor (1-\delta)N \rfloor - {l/2}$, for $0< \delta <1$, are classically simulatable.

\section{Quantum circuit representation of a 2D self-dual kicked Ising model  \label{sec:appC}} 

In this appendix, we show that a 2D self-dual kicked Ising {model} is represented as a 2D DUQC. Let us consider a  2D self-dual kicked Ising {model}, associated with a 2N $\times$ 2N square lattice:
\begin{eqnarray}
&H_{\rm{2DKI}}(t)=H_{\rm{I}}+\sum_{n=-\infty}^{\infty}\delta (t-n)H_{\rm{K}}, \label{2DKI}\\
&H_{\rm{I}}=\sum_{j,k=1}^{2N} \left\{ J\left( Z_{j,k}Z_{j+1,k}+Z_{j,k}Z_{j,k+1} \right)+hZ_{j,k} \right\},\\
&H_{\rm{K}}=b\sum_{j,k=1}^{2N}X_{j,k},
\end{eqnarray}
where $\delta (t)$ is the Dirac delta function, {$|J|$ and $|b|$ are equal to $\frac{\pi}{4}$}, $h$ is an arbitrary real number, and we adopted {PBCs}, {that is,} $Z_{j,2N+1}=Z_{j,1}$ and $Z_{2N+1,k}=Z_{1,k}$. The Floquet operator associated to Eq. (\ref{2DKI}) can be written as
\begin{equation}
    \mathcal{U}_{\rm{KI}}\coloneqq \mathcal{T}e^{-\int_0^1dtH(t)}=\textbf{U}_{\rm{K}} \textbf{U}_{\rm{I1}}\textbf{U}_{\rm{I2}}\textbf{U}_{\rm{I3}}\textbf{U}_{\rm{I4}},
    \end{equation}
    where $\mathcal{T}$ denotes a time ordered product, and we define $\textbf{U}_{\rm{K}}$, $\textbf{U}_{\rm{I1}}$, $\textbf{U}_{\rm{I2}}$, $\textbf{U}_{\rm{I3}}$, and  $\textbf{U}_{\rm{I4}}$ as follows:
    
    \begin{gather}
    \textbf{U}_{\rm{K}}\coloneqq e^{-\sum_{j,k=1}^{2N}bX_{j,k}},
    \\
    \textbf{U}_{\rm{I1}}\coloneqq e^{-\sum_{j=1}^{N}\sum_{k=1}^{2N} \left( JZ_{2j,k}Z_{2j+1,k}+hZ_{2j,k}\right)},\\
    \textbf{U}_{\rm{I2}}\coloneqq e^{-\sum_{j=1}^{2N}\sum_{k=1}^{N}JZ_{j,2k-1}Z_{j,2k}}, \label{UI2} \\
    \textbf{U}_{\rm{I3}}\coloneqq e^{-\sum_{j=1}^{N}\sum_{k=1}^{2N} \left(JZ_{2j-1,k}Z_{2j,k}+hZ_{2j-1,k} \right)},\\ 
    \textbf{U}_{\rm{I4}}\coloneqq e^{-\sum_{j=1}^{2N}\sum_{k=1}^{N}JZ_{j,2k}Z_{j,2k+1}}.\label{UI4}
\end{gather}
Then, the integer powers of the Floquet operator have forms:
\begin{gather}
\mathcal{U}_{\rm{KI}}^{2t}=\textbf{U}_{\rm{K}}\textbf{U}_{\rm{I3}}\textbf{U}_{\rm{I2}}\textbf{U}_{\rm{I4}}\textbf{U}_{\rm{KI1}}(\textbf{U}_{\rm{I2}}\textbf{U}_{\rm{I4}}\textbf{U}_{\rm{KI3}}\textbf{U}_{\rm{I2}}\textbf{U}_{\rm{I4}}\textbf{U}_{\rm{KI1}})^{t-1}\textbf{U}_{\rm{I2}}\textbf{U}_{\rm{I4}}\textbf{U}_{\rm{I3}},\\
\mathcal{U}_{\rm{KI}}^{2t+1}=\textbf{U}_{\rm{K}}\textbf{U}_{\rm{I1}}(\textbf{U}_{\rm{I2}}\textbf{U}_{\rm{I4}}\textbf{U}_{\rm{KI3}}\textbf{U}_{\rm{I2}}\textbf{U}_{\rm{I4}}\textbf{U}_{\rm{KI1}})^{t}\textbf{U}_{\rm{I2}}\textbf{U}_{\rm{I4}}\textbf{U}_{\rm{I3}},
\end{gather}
where we define
\begin{gather}
\textbf{U}_{\rm{KI1}}=\textbf{U}_{\rm{I1}}\textbf{U}_{\rm{K}}\textbf{U}_{\rm{I1}},\\
\textbf{U}_{\rm{KI3}}=\textbf{U}_{\rm{I3}}\textbf{U}_{\rm{K}}\textbf{U}_{\rm{I3}}.
\end{gather}
Using the fact that $\textbf{U}_{\rm{KI1}}$ and $\textbf{U}_{\rm{KI3}}$ are written as dual-unitary gates \cite{PhysRevLett.123.210601}, it {follows} that the quantum circuit $(\textbf{U}_{\rm{I2}}\textbf{U}_{\rm{I4}}\textbf{U}_{\rm{KI3}}\textbf{U}_{\rm{I2}}\textbf{U}_{\rm{I4}}\textbf{U}_{\rm{KI1}})^{t}$ is one of 2D DUQCs. {We note that this is even true if interaction strength $J$ in Eqs. (\ref{UI2}) and (\ref{UI4}) are replaced by arbitrary real numbers. This is because unitary gates of 2D DUQCs in $k$-direction can be chosen arbitrarily.}

{
\section{Local observables of DUQCs with open boundary conditions \label{sec:appD}} 
In this appendix, we discuss classical simulatability of DUQCs under OBCs.
First, we show that local expectation values of 1D DUQCs with OBCs at an early time become dependent on dual-unitary gates but still classically simulatable. 
We define 1D DUQCs on 2N qubits with OBCs as the following:
\begin{equation}
\bm{V}(t) = 
\includegraphics[clip,width=3cm,valign=c]{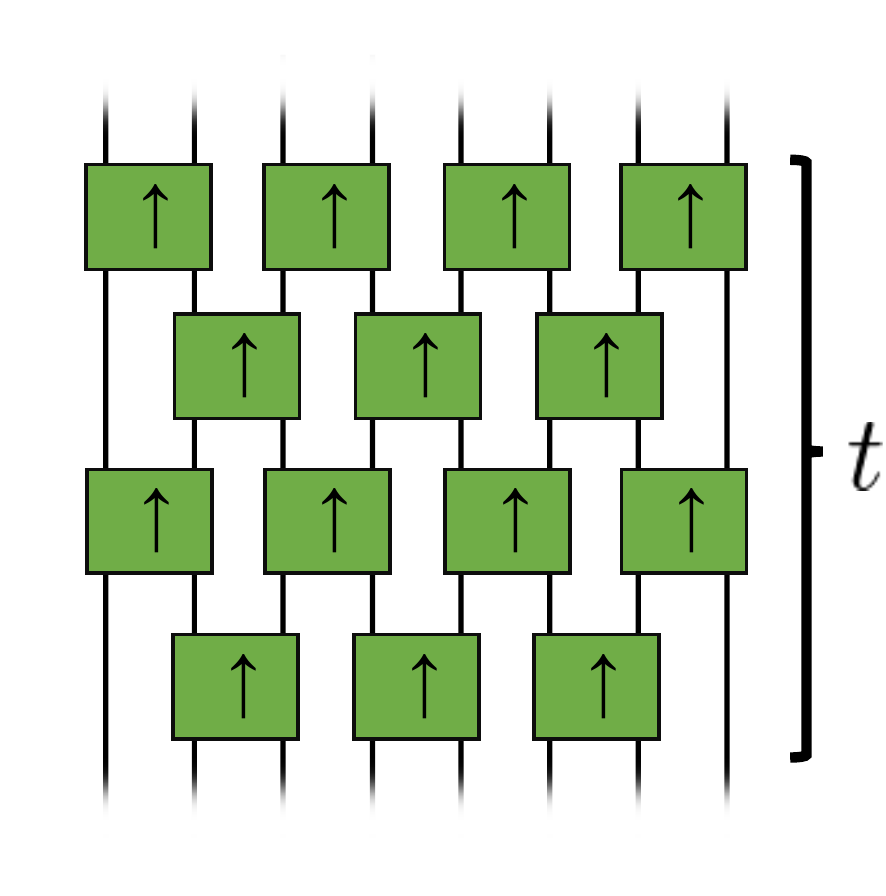}. \label{fig.1Dqc_OBC}
\end{equation}
Solvable initial states with OBCs are defined as with Eq. (\ref{eq:alphabeta}).
Then, expectation values of local observables $O$, after removing dual-unitary gates outside of the causal-cone, can be written as 
\begin{equation} \label{E1_OBC}
\braket{\Psi_{(\alpha^*, \beta^*)}|\bm{V}(t)^{\dagger}O\bm{V}(t)|\Psi_{(\alpha, \beta)}}=
\includegraphics[clip,width=6.5cm,valign=c]{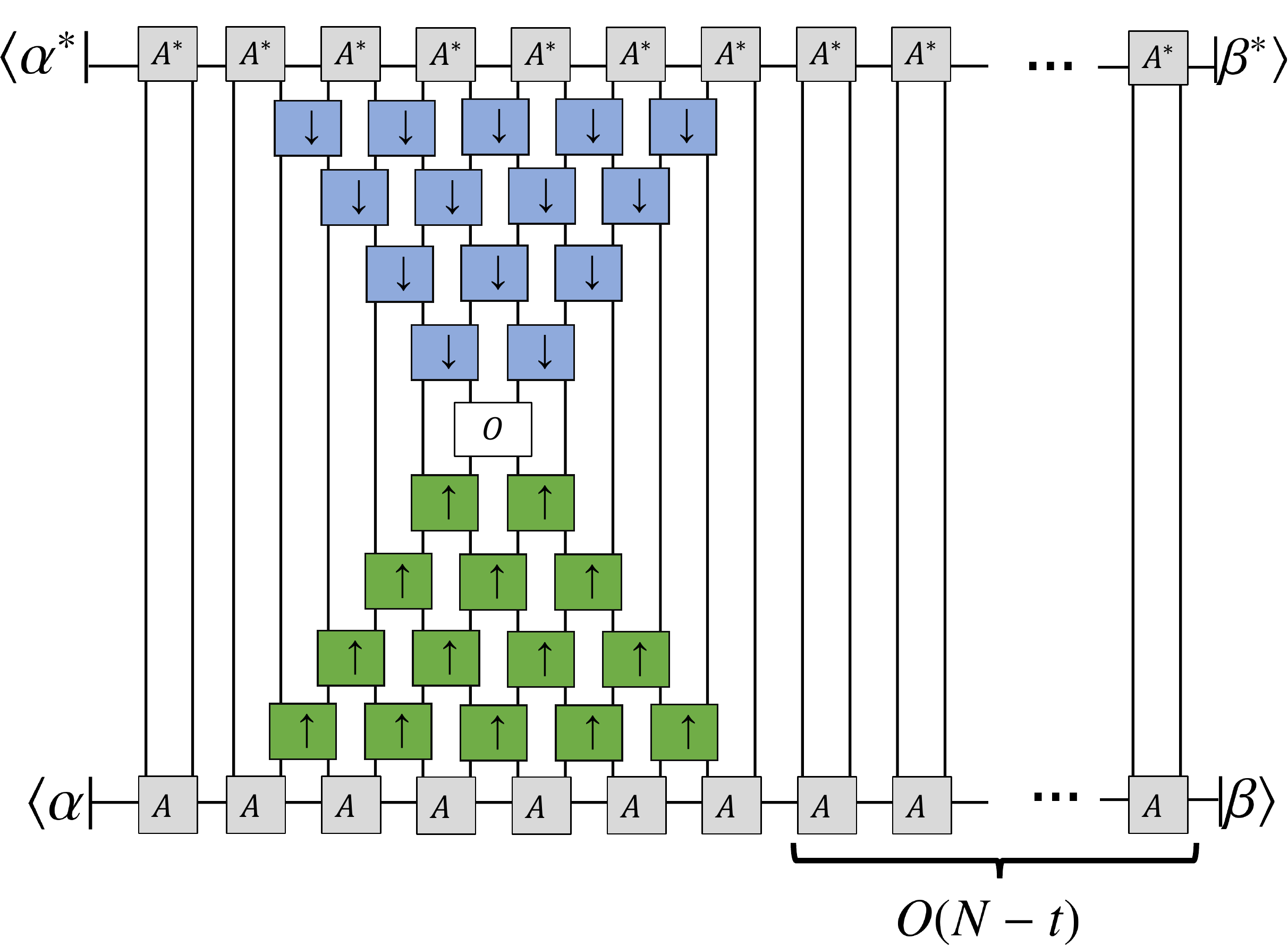},
\end{equation}
where without loss of generality we assume that a local observable is supported on the left half of chain.
From the argument in Appendix \ref{sec:appC} and by using Eqs. (\ref{ducont}) and (\ref{matrix_unitarity}), it can be easily shown that Eq. (\ref{E1_OBC}) is exponentially close to
\begin{equation}
\includegraphics[clip,width=3cm,valign=c]{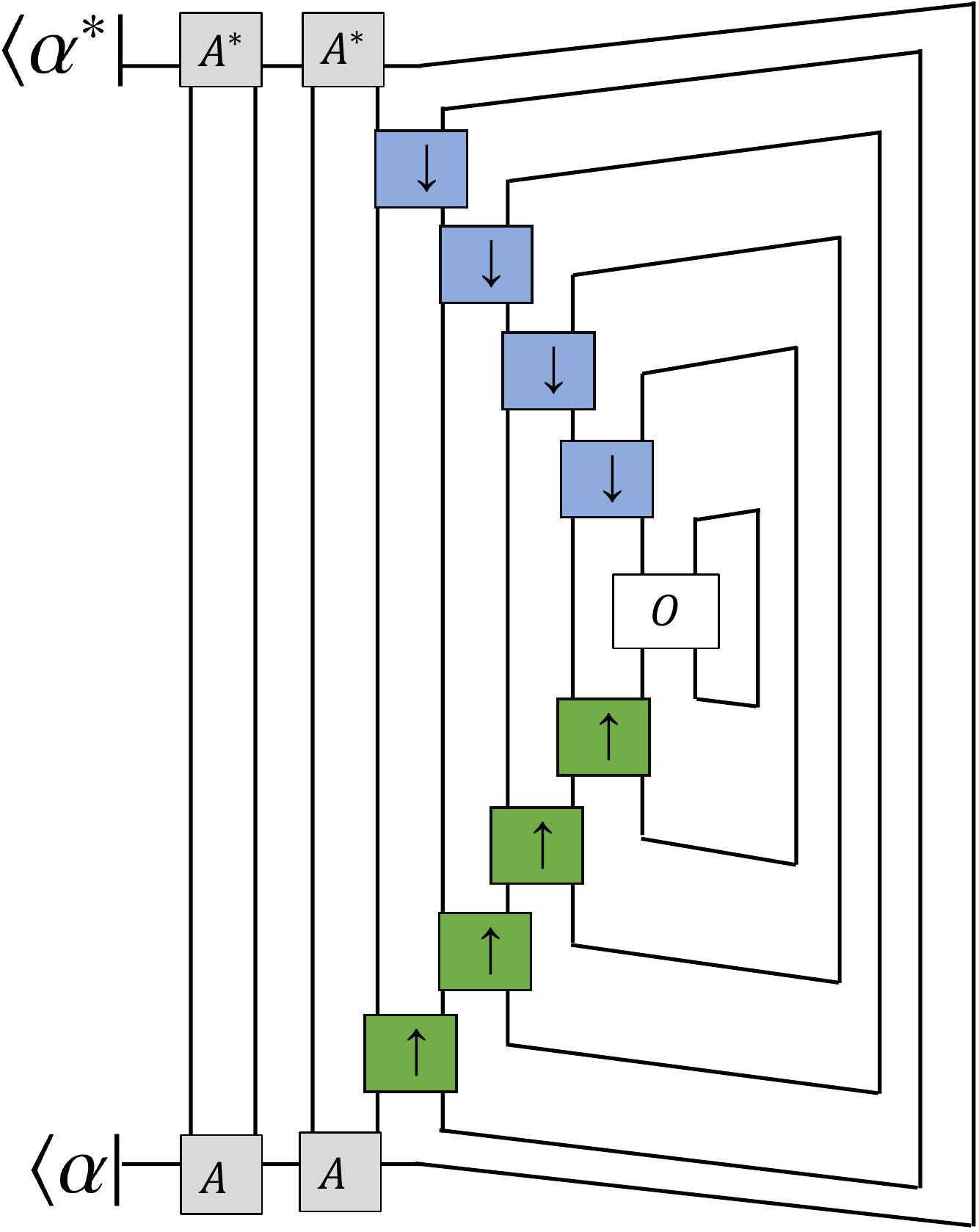}.
\end{equation}
It is dependent only on dual-unitary gates on the boundary of the causal-cone, and therefore it is classically simulatable.
}

{
However, when we generalize the above argument to two-dimensional cases, local expectation values do not seem to be classically simulatable in linear depth because uncontracted unitary gates on the boundary of the causal-cone form 2D tensor-networks.
This situation is similar to that of correlation functions for 2D DUQCs discussed in Sec. \ref{subsec_Correlation functions} of the main text.
Besides, it is reminiscent of matchgate circuits, where classical simulatability depends on their connectivity \cite{brod2012geometries}. 
As written in Sec. \ref{sec:Conclusion}, it would be interesting future work to characterize the computational power of DUQCs with various connectivity.
}

\bibliographystyle{unsrturl}
\bibliography{quantum-template}% Produces the bibliography via BibTeX.

\end{document}